\documentclass[3p,authoryear,a4paper]{elsarticle}
\usepackage{graphicx}
\usepackage{slashbox}
\usepackage{color}
\usepackage{dsfont}
\usepackage{amsmath}
\usepackage{amssymb}
\usepackage{amsfonts}
\usepackage{amsbsy}
\usepackage{caption}
\usepackage{subcaption}
\usepackage{amsthm}
\usepackage{array}
\usepackage{booktabs} 
\usepackage{verbatim}
\usepackage[english]{babel}
\usepackage{natbib}
\usepackage{ulem}
\usepackage{url}
\usepackage{multirow}
\usepackage{float}
\usepackage{numprint}
\usepackage{enumitem}
\usepackage[title]{appendix}
\usepackage[font=normalsize, labelfont=bf]{caption}
\usepackage{tabularx}
\usepackage{ltablex}
\usepackage{makecell}
\usepackage[table,xcdraw]{xcolor}
\usepackage{longtable}
\usepackage{hyperref}
\usepackage{listings}

\lstdefinestyle{codelisting}{
    basicstyle=\ttfamily\small,
    breakatwhitespace=false,         
    breaklines=true,                 
    captionpos=b,                    
    keepspaces=true,                 
    numbers=left,                    
    numbersep=5pt,                  
    showspaces=false,                
    showstringspaces=false,
    showtabs=false,                  
    tabsize=2
}

\numberwithin{table}{section}

\newcolumntype{R}{>{\raggedleft\arraybackslash}X} 
\newcolumntype{P}[1]{>{\centering\arraybackslash}p{#1}} 

\renewcommand\appendix{\par
\setcounter{section}{0}
\setcounter{subsection}{0}
\setcounter{table}{0}
\setcounter{figure}{0}
\gdef\thetable{\Alph{table}}
\gdef\thefigure{\Alph{figure}}
\gdef\thesection{\Alph{section}}
\setcounter{section}{0}}

\newtheorem{theorem}{Theorem}

\newtheorem{principle}[theorem]{Principle} 
\numberwithin{equation}{section}

\newenvironment{arcitem}{
\begin{list}{---}{
\topsep=1pt
\itemsep=1pt 
\parsep=0pt 
\leftmargin=19pt 
}}
{\end{list}}

\newcounter{arclist}

\newcounter{arcenum}

\newcommand{\rev}[1]{\textcolor{black}{#1}}

\begin{document}

\normalem

\begin{frontmatter}

\title{\texttt{SPLICE}: A Synthetic Paid Loss and Incurred Cost Experience Simulator}

\author[UMelb]{Benjamin Avanzi}
\ead{b.avanzi@unimelb.edu.au}

\author[UNSW]{Greg Taylor\corref{cor}}
\ead{gregory.taylor@unsw.edu.au}

\author[UNSW]{Melantha Wang}
\ead{chenyi.wang@student.unsw.edu.au}


\cortext[cor]{Corresponding author.}

\address[UMelb]{Centre for Actuarial Studies, Department of Economics, University of Melbourne VIC 3010, Australia}
\address[UNSW]{School of Risk and Actuarial Studies, UNSW Australia Business School, UNSW Sydney NSW 2052, Australia}

\begin{abstract}
In this paper, we first introduce a simulator of cases estimates of \textit{incurred losses}, called \texttt{SPLICE} (\texttt{S}ynthetic \texttt{P}aid \texttt{L}oss and \texttt{I}ncurred \texttt{C}ost \texttt{E}xperience). In three modules, case estimates are simulated in continuous time, and a record is output for each individual claim. Revisions for the case estimates are also simulated as a sequence over the lifetime of the claim, in a number of different situations. Furthermore, some dependencies in relation to case estimates of incurred losses are incorporated, particularly recognizing certain properties of case estimates that are found in practice.  For example, the magnitude of revisions depends on ultimate claim size, as does the distribution of the revisions over time. Some of these revisions occur in response to occurrence of claim payments, and so \texttt{SPLICE} requires input of simulated per-claim payment histories. The claim data can be summarized by accident and payment ``periods'' whose duration is an arbitrary choice (e.g. month, quarter, etc.) available to the user. 

\texttt{SPLICE} is built on an existing simulator of individual claim experience called \texttt{SynthETIC} \citep*[introduced in][]{AvTaWaWo21,R-SynthETIC}, which offers flexible modelling of occurrence, notification, as well as the timing and magnitude of individual partial \textit{payments}. This is in contrast with the \textit{incurred losses}, which constitute the additional contribution of \texttt{SPLICE}. The inclusion of incurred loss estimates provides a facility that almost no other simulators do.


\texttt{SPLICE} is is a fully documented R package that is publicly available and open source (on CRAN). \texttt{SPLICE}, combined with \texttt{SynthETIC}, provides eleven modules (occurrence, notification, etc.), any one or more of which may be re-designed according to the user’s requirements. It comes with a default version that is loosely calibrated to resemble a specific (but anonymous) Auto Bodily Injury portfolio\rev{, as well as data generation functionality that outputs alternative data sets under a range of hypothetical scenarios differing in complexity}.  The general structure is suitable for most lines of business, with some re-parameterization.

\end{abstract}

\begin{keyword} 
case estimates, granular models, incurred loss, individual claims, individual claim simulator, loss reserving, partial payments, simulated losses, superimposed inflation, SynthETIC, synthetic losses.

JEL codes: 
C52 \sep  
C53 \sep  
C63 \sep  
G22 

MSC classes:
91G70 \sep 	
91G60 \sep 	
62P05 

\end{keyword}
\end{frontmatter}
{\centering \large}

\section{Introduction}\label{sec:intro}

\subsection{Background and motivation} \label{ssec:background}

Machine learning methods are developing at increasing speeds in the actuarial literature. While the ultimate objective of these machine learning methods is application to real data, availability of synthetic data containing features commonly observed in real data is useful for at least two reasons: (i) such data sets, especially of granular nature and of large size, are in short supply in the actuarial literature \citep*[see, e.g., Section 2.3 of][]{EmWu21}, (ii) knowledge of the data generating process (impossible with real data) assists with the validation of the strengths and weaknesses of any new methodology.

Referring to scarcity of data (item (i) above), \citet*[Section 2.3]{EmWu21} mention two stochastic scenario generators: \citet*{GaWu18}, and \citet*{AvTaWaWo21,R-SynthETIC}. These simulators \citep*[and others, see for instance Section 5 of][for a comprehensive review]{AvTaWaWo21} produce synthetic claims \emph{payments} experience. With the exception of the simulator developed by the ``ASTIN Working Party on Individual Claim Development with Machine Learning'' \citep*{HaGaJa17}, none of those simulators consider \textbf{case estimates of incurred losses}; see also Section~\ref{sec:prior lit} for further detail. These can be of significance for inference and prediction. An earlier reference on that point is \citet*[Volume 3, p. 1383 et seq.]{TaSu04}; see also \citet*{MaQu04} on Munich Chain Ladder, and  \citet*{TaMcSu08}, who exemplify improved forecast performance when case estimates are taken into account.

The present paper describes an extension of \texttt{SynthETIC}, called \textbf{\texttt{SPLICE} (\texttt{S}ynthetic \texttt{P}aid \texttt{L}oss and \texttt{I}ncurred \texttt{C}ost \texttt{E}xperience)}, whose purpose is to close this gap in a manner consistent with the earlier paid claim experience, while leaving feature control on the hands of the user. In three modules, case estimates are simulated in continuous time, and a record is output for each individual claim. Revisions for the case estimates are also simulated as a sequence over the lifetime of the claim, in a number of different situations. Some of these revisions occur in response to occurrence of claim payments, and so \texttt{SPLICE} requires input of simulated per-claim payment histories. Furthermore, some dependencies in relation to case estimates of incurred losses are incorporated, particularly recognizing certain properties of case estimates that are found in practice; a full list of our modelling overarching principles is provided in Section \ref{ssec:case estimates descript} and Section \ref{S_datafeatures}.  For example, the magnitude of revisions depends on ultimate claim size, as does the distribution of the revisions over time. The claim data can be summarized by accident and payment ``periods'' whose duration is an arbitrary choice (e.g. month, quarter, etc.) available to the user.

Although the three additional incurred loss modules in \texttt{SPLICE} could theoretically be used with any claims occurrence, notification and payment base (with care), we chose to make \texttt{SynthETIC} a required package, and use its existing structure without modifications. \texttt{SynthETIC} simulates paid claim experience of individual claims at a transactional level (key dates associated with a claim---e.g. settlement date---and claim payments). It offers an extremely flexible and complex base already: its modules represent specific features of the paid claim experience (e.g. claim sizes), and their plug-in nature hands control of these features to the user. Furthermore, when considering the output of \texttt{SynthETIC} along with those of \texttt{SPLICE}, the transactional simulation output now comprises key dates, and both claim payments and revisions of estimated incurred losses, and all this in a coherent and flexible format. This justifies the name of the package introduced in this paper: \texttt{S}ynthetic \texttt{P}aid \texttt{L}oss and \texttt{I}ncurred \texttt{C}ost \texttt{E}xperience (\texttt{SPLICE}).

Finally, we refer to reason (ii) above, which mentioned the usefulness of synthetic data for model development and validation; see also Section \ref{S_gran}. The \textit{Annals of Actuarial Science} requires that submissions to its Actuarial Software stream demonstrate the analysis workflow using both synthetic and real data, highlighting the importance of synthetic data in this regard. When using \texttt{SynthETIC} and \texttt{SPLICE} together, the user has full control of the mechanics of the evolution of an individual claim. In particular, the user can decide the level of dependencies to include between different claim variates and test the effectiveness of any proposed new model in detecting such interactions. Indeed, by testing the proposed model against data across a spectrum of complexity, the user may derive new insights into the its strengths and weaknesses, which is one main advantage of using synthetic data over real data. \rev{This is developed in Section \ref{ssec:alternative-datasets}.}

\subsection{Relation to prior literature}\label{sec:prior lit}

\subsubsection{Claim simulation literature}\label{ssec: simulation lit}

\citet*{AvTaWaWo21} discussed a few predecessor simulators to some detail: \citet*{HaGaJa17,GaWu18,  Co20, CAS07, CAS11, R-cascsim}. Of these, only the last three have been shown to simulate case estimates.  All three generate samples of both paid and incurred amounts. 

\citet*{CAS07} was superseded by the more advanced CAS simulator \citep*{CAS11}.  Here, case estimates are generated over the lifetime of each claim in continuous time.

Dates of incurred loss revisions are assigned randomly and independently over the claim lifetime according to a specified distribution.  In most cases, a revision at (continuous) development time $j$ is generated in the form
$$ \text{Estimate of incurred loss}=\text{ultimate claim size}\times\text{adequacy factor},$$
where the adequacy factor is drawn from a log normal distribution.  The parameters of this distribution may be defined by the user at fractions 0\%, 40\%, 70\% and 90\% of the claim lifetime.  For intermediate fractions, the simulator interpolates the log normal mean.  Drawings of distinct revisions appear to be stochastically independent.

This last feature ensures that the magnitudes of the revisions over the claim lifetime can be controlled.  However, these revisions are independent of the size of the claim itself.  \texttt{SPLICE} incorporates dependency in this respect.  It also distinguishes between major and minor revisions, with further dependencies between the magnitudes of multiple revisions in respect of the same claim.

\citet*{CAS11} allows for inclusion of inflation according to accident period and, optionally, calendar period. In the latter case, the rates of accident and calendar period inflation are related. \texttt{SPLICE} allows the specification of arbitrary rates of inflation of either type. Calendar period inflation is specified as base inflation plus a superimposed inflation component. The rate of superimposed inflation may vary from claim to claim according to claim attributes such as ultimate size.

The more recent claim simulator sponsored by the Casualty Actuarial Society \citep{R-cascsim} is structured differently from \citet*{CAS11}.  It appears to be concerned more with the simulation of the ultimate individual costs of a given portfolio of claims than with simulation of the detailed development of each claim.

It contains four options for the simulation of ultimate incurred loss, but only one of these generates a series of case estimates over the life of the claim. The other three simulate just ultimate incurred cost from the current claim status.

The one option that does simulate the development of incurred cost over claim lifetime does so by means of year-to-year development factors that are sampled from distributions defined by the user. These distributions differ from one development year to another, but the sampled development factors are stochastically independent, and there is no apparent provision for them to depend on the existing claim status (e.g. total paid to date). \texttt{SPLICE} remedies this.

In summary, \texttt{SPLICE} provides the following enhancements:
\begin{arcitem}
\item Major and minor revisions are differentiated.
\item The frequencies and magnitudes of these revisions can be made to depend on claim attributes such as ultimate cost.
\item Dependencies are introduced between the magnitudes of revisions in respect of a single claim.
\item The forms of inflation included are very general and flexible.
\item Distributions of frequency and severity of revisions can be specified in a flexible way (a quality inherited from \texttt{SynthETIC}), that is, beyond lognormal. The complexity allowed here flows on to the incurred losses through their dependence on payment amounts and ultimate cost.
\end{arcitem}

A brief mention of the rather different simulator of \citet*{Co20} is appropriate here.  This deals with a given set of data points of unspecified form, which might therefore, in principle, comprise time series of payments and case estimates for each claim.

The purpose of the suggested algorithm is to generate a synthetic data set with the same stochastic properties as the original. Generative adversarial networks are used to infer an underlying distribution of the data points, and then a new sample is drawn from this distribution.

The process is exemplified in \citet*{Co20} using a well known portfolio of French motor third-party liability policies \citep*[from \texttt{CASdatasets},][]{R-CASdatasets}. The individual claim portfolio is re-sampled with respect to various claim attributes (car age, driver age, etc.) and claim count. However, no example demonstrating a re-sampling of paid claims and incurred claim costs is yet available.

\subsubsection{Granular model literature} \label{S_gran}
The development of  so-called  ``granular (or micro-) models'' requires availability of data at a certain level of detail.  As such, such development is likely to benefit from the availability of simulator such as \texttt{SPLICE}, which generates this detail. 
\citet*{DeMo19} provides a summary of the literature of these models, which are also discussed by \citet*{Tay19}.  However, the authors are unaware of any contributions to the granular model literature that consider the evolution of case estimates.

\subsection{Package installation} \label{sec:SPLICE repo}
\texttt{SPLICE} is released as an open-source R package on the Comprehensive R Archive Network (CRAN) at \url{https://CRAN.R-project.org/package=SPLICE} \citep*{R-SPLICE}. In combination with \texttt{SynthETIC}, an existing open-source simulator of paid losses of individual claims available on CRAN \citep*{R-SynthETIC}, \texttt{SPLICE} simulates sequentially each of the eleven modules as outlined in Section~\ref{sec:architecture}, which provide the full functionality for the generation of synthetic paid loss and incurred cost experience (see Section~\ref{sec:architecture}). Its modular structure is discussed in Section~\ref{ssec:architecture claim process}.

\texttt{SPLICE} offers a collection of simulation functions for the incurred estimates. The default parameters for the simulation of revisions of incurred loss estimates are detailed in Appendix~\ref{appendix:A}, but they can be easily modified (or unplugged and replaced, if needed) by users to match their own experience with case estimates.

Users can choose to output their simulated claim histories in the form of a chain-ladder square of incurred losses by occurrence and development periods, or an individual transactional data set. In the latter, a transaction can be any of claim notification, settlement, a payment, or a case estimate revision. A test transactional data set generated under the current specification is also available as part of the package (an example data excerpt is included in Appendix~\ref{appendix:data excerpt}). A full demonstration of the features offered by \texttt{SPLICE} can be accessed by running \texttt{vignette("SPLICE-demo", package = "SPLICE")} in the R console after the installation of the package. 

Users can install the latest version of \texttt{SPLICE} from the CRAN repository via

\texttt{> install.packages("SPLICE")}

A development version of the program is also available on \url{https://github.com/agi-lab/SPLICE}. The GitHub repository contains, in addition to the package code, a chain-ladder analysis of the test data set discussed in Section~\ref{ssec:example-implementation}, in an Excel spreadsheet, \rev{as well as some example datasets of various levels of complexity described in Section~\ref{ssec:alternative-datasets}, available for download in \texttt{.csv} format.}

\subsection{Structure of the paper}

The claim process is defined by the original 8 paid loss modules (from \texttt{SynthETIC}) and an additional 3 incurred loss modules: major revisions, minor revisions, and consolidation of revisions, with the option to include inflation; see Section~\ref{sec:architecture}. \texttt{SPLICE} allows the user to specify any alternative form of distribution for the simulation of frequency, timing and magnitude of incurred loss revisions. The current default parameterization has been set up to resemble the evolution of case estimates commonly found in practice; see also Section \ref{S_datafeatures}.

The general nature of case estimates is described in Section~\ref{sec:case estimates}.  After some notation in Section~\ref{sec:notation}, \texttt{SPLICE} architecture is described in Section~\ref{sec:architecture}.  
\rev{Section~\ref{sec:SPLICE-application} demonstrates the application of \texttt{SPLICE}, including an example implementation with the default parameterization just mentioned, as well as an illustration of how alternative scenarios can be generated to achieve different levels of complexity.} Section~\ref{sec:conclusion} contains some closing comments.

\section{Case Estimates}\label{sec:case estimates}
\subsection{General description}\label{ssec:case estimates descript}

Most insurers assign \textbf{case estimates} to individual claims.  A case estimate is here defined to mean an estimate of the ultimate cost of a claim, arrived at subjectively by means of expert knowledge.  Case estimates are sometimes referred to as \textbf{manual estimates} or \textbf{physical estimates}.  The experts who formulate them are usually known as \textbf{case estimators} or \textbf{loss adjusters}.

It is assumed that each claim carries, at each point in its lifetime, a case estimate of its ultimate incurred cost, and that the case estimators will vary these over time as additional information comes to hand.

It is assumed that revisions are either \textbf{major} or \textbf{minor}.  Major revisions occur in response to material new evidence.  For example, a claimant suffering head injury may be medically declared vegetative, in which case the perceived claim liability might increase substantially.  Minor revisions occur as a result of more routine vagaries of a claim’s progress.  For example, unforeseen medical reports might be required.

Major revisions will be infrequent and usually of greater magnitude than minor.  Moreover, major revisions represent a total change of perspective on ultimate claim cost, causing the case estimator to apply a revision factor to his estimate of that cost.  Minor revisions, on the other hand, respond more to matters of detail, causing the case estimator to apply a revision factor to his estimate of outstanding payments.

The points below describe the development of a case estimate over the lifetime of a claim.  Practice varies from one insurer to another, and the description given here may not fit all insurers.  It would, however, describe a common practice.

The case estimate relates to the ultimate cost of the claim.  Since the outstanding amount of the claim is equal to the difference between the ultimate cost and the paid losses to date, and since the latter is known at any point of the claim’s lifetime, it follows that a case estimate of ultimate cost implies a case estimate of outstanding amount and vice versa.

The assumed features of incurred claims included in \texttt{SPLICE} are guided by the following overarching realistic principles:

\begin{principle}
The insurer maintains case estimates of the incurred loss, and hence the outstanding loss, associated with each notified claim.\label{principles: principle1}
\end{principle}
\begin{principle}As long as there is no revision of the incurred loss, the estimate of outstanding loss is written down by each partial payment as it is made.  This process is automated, and there is no intervention by the case estimator.\label{principles: principle2}
\end{principle}
\begin{principle}The case estimate of the incurred loss may undergo a number of revisions over the claim’s lifetime.\label{principles: principle3}
\end{principle}
\begin{principle}These may occur at the time of a partial payment, or at any other time.\label{principles: principle4}
\end{principle}
\begin{principle}These revisions may be major (e.g. increase by a factor of 5) or minor (e.g. decrease by 5\%).\label{principles: principle5}
\end{principle}
\begin{principle}By convention, each claim undergoes its first major revision at notification, when a case estimate is first established.\label{principles: principle6}
\end{principle}
\begin{principle}Major revisions other than this initial one are more likely for larger claims, and do not occur at all for the smallest claims.\label{principles: principle7}
\end{principle}
\begin{principle}They are relatively unlikely in the latter part of the claim’s lifetime.\label{principles: principle8}
\end{principle}
\begin{principle}A claim may experience up to two major revisions in addition to the initial one, but the second, if it occurs at all, is likely to be smaller than the first.\label{principles: principle9}
\end{principle}
\begin{principle}Minor revisions tend to be upward in the early part of a claim’s life, and downward in the latter part.\label{principles: principle10}
\end{principle}
\begin{principle}At settlement of the claim, the case estimate of ultimate cost will, by principle, be equal to actual amount paid, adjusted to the settlement date for base inflation.  Correspondingly, the case estimate of outstanding claim cost will, by principle, be equal to zero.\label{principles: principle11}
\end{principle}

\rev{Although these are all listed as ``principles'', it would also be fair to regard Principles \ref{principles: principle7}--\ref{principles: principle10} as \textbf{design features}. They are elevated to the status of ``principles'' here because they are, at least in the cases of ~\ref{principles: principle7}--\ref{principles: principle9}, commonly observed in practical claim portfolios. Principle \ref{principles: principle10} is somewhat different. Although it is encountered in many portfolios, alternatives are often encountered. The user should bear in mind that Principles \ref{principles: principle7}--\ref{principles: principle10} are, nonetheless, discretionary features of the default version of the simulator, and there is ample scope for their variation. For example, the user may choose to allow more than three major revisions over the course of the claim and make respective changes to the simulation of revision multipliers (which, by default, assumes a maximum of three major revisions).}

\subsection{Treatment of inflation}\label{ssec: treatment of inflation}
Here, \textbf{base inflation} is defined in Section~\ref{ssec: notation claim payments} below as ``normal'' community inflation, such as price inflation or wage inflation, that would apply to claim sizes in the absence of extraordinary considerations.  It is to be contrasted with \textbf{superimposed inflation (``SI”)}, which represents the difference between the total rate of escalation of claim costs and base inflation. These principles are consistent with those of the original \texttt{SynthETIC} \citep*{R-SynthETIC}, but some additional assumptions and explanations are required with respect to the (new) incurred claims component.

Typically, case estimators are not expected to anticipate future base inflation.  They may often be requested explicitly to exclude inflation beyond the valuation date.  In such a system, each case estimate will represent ultimate claim cost in current-day values.  This approach allows the insurance management to incorporate its own assumptions for future base inflation, which may depend on within-insurer consensus on economic conditions.

The estimators will usually be required to include full superimposed inflation up to the date of claim settlement.

Between case estimate revisions (see Principles~\ref{principles: principle2}--\ref{principles: principle10}), the estimated ultimate claim cost remains unchanged.  Partial payments may occur, and the case estimate of outstanding claims will respond (see Principle~\ref{principles: principle2}), but the estimate of ultimate cost remains unchanged.

Revisions of incurred cost are the only points of intervention of the case estimators. At any such point, the estimator will adjust for base inflation to that point, i.e. an adjustment for the time elapsed since the immediately preceding revision.  A ``current-day value revision factor” will then be applied, representing the estimator’s change of opinion in ultimate claim size but with no allowance for any base inflation beyond the date of valuation.

\section{Notation}\label{sec:notation}
\subsection{Claim payments}\label{ssec: notation claim payments}
The notation for claim payments was set out in \citet*{AvTaWaWo21}. It is repeated here in Section \ref{ssec: notation claim payments} almost verbatim for convenience, as some will be required for the description of the incurred loss simulation.  Its repetition will also provide a comprehensive view of the simulator in its entirety. 

\texttt{SPLICE} works with \textbf{exact transaction times}, so time will be measured continuously.  Calendar time $\bar{t}=0$ denotes the first date on which there is exposure to occurrence of a claim.  The time scale is arbitrary; a unit of time might be a quarter, a year, or any other selected period.  The length of a period in years is specified by the user as a global parameter.  The user needs to ensure that all input parameters are compatible with the chosen time unit.  

For certain purposes (see Section~\ref{sec:architecture}), it will be useful to partition time into discrete periods.  These are unit periods according to the chosen time scale.  These periods will be of two types:
\begin{arcitem}
    \item \textbf{occurrence periods} (or accident periods), numbered $1,2, \dots, I$, where occurrence period $1$ corresponds to the calendar time interval $(0,1]$;
    \item \textbf{payment periods}, numbered $1,2, \dots, 2I-1$, representing the calendar periods in which individual payments are made, and including I past periods and a further $I-1$ future ones.
\end{arcitem}

An individual claim is settled by means of one or more separate payments, referred to here as \textbf{partial payments}.  The claim will be regarded as settled immediately after the final partial payment.  The delays between successive partial payments are referred to as \textbf{inter-partial delays}.

All payments are subject to inflation.  They are initially simulated without allowance for inflation, and an inflation adjustment added subsequently.  Any quantity described as ``without allowance for inflation” is expressed in constant dollar values, specifically those of payment period $1$. Inflation occurs in two types:
\begin{enumerate}[label=(\alph*)]
    \item \textbf{Base inflation}: which represents, in some sense, ``normal” community inflation (e.g. price inflation, wage inflation) that would apply to claim sizes in the absence of extraordinary considerations; and
    \item \textbf{Superimposed inflation}: which represents the differential (positive or negative) between claim inflation and base inflation.
\end{enumerate}
It is assumed that base inflation may be represented by a vector of quarterly inflation rates for both past and future calendar periods.  The input inflation rates need to be expressed as quarterly effective rates irrespective of the length of calendar periods adopted.  The inflation rates are used to construct an inflation index whose values are obtained:
\begin{arcitem}
    \item at quarterly points from calendar time 0, by compounding the quarterly rates; and
    \item at intra-quarterly points, by exponential interpolation between the quarter ends immediately prior and subsequent.
\end{arcitem}
It is also assumed that SI occurs in two sub-types:
\begin{enumerate}[label=(\roman*)]
    \item \textbf{Payment period SI}: which operates over payment periods; and
    \item \textbf{Occurrence period SI}: which operates over occurrence periods.
\end{enumerate}
The following notation is used throughout:
\begin{arcitem}
    \item[$\centerdot$] $\lfloor x \rfloor$ denotes the integral part of $x$
    \item[$\centerdot$] $\lceil x \rceil = $ the ceiling function $\lceil x \rceil$ = integral $n$ for $n-1 < x \leq n$
    \item[$\centerdot$] $i =$ occurrence period $1,2,\dots, I$
    \item[$\centerdot$] $\bar{t} =$ continuous calendar time with origin at the beginning of occurrence period 1
    \item[$\centerdot$] $t = \lceil \bar{t} \rceil =$ payment period
    \item[$\centerdot$] $E_i =$ (annual effective) exposure in occurrence period $i$
    \item[$\centerdot$] $\lambda_i =$ expected claim frequency (per unit exposure) in occurrence period $i$
    \item[$\centerdot$] $f(\bar{t}) =$ base inflation index, representing the ratio of dollar values at calendar time $\bar{t}$ to those at calendar time 0, constructed from the input base inflation rates
    \item[$\centerdot$] $g_P (\bar{t}|s) =$ payment period SI index, representing the ratio of dollar values at calendar time $\bar{t}$ to those at calendar time 0
    \item[$\centerdot$] $g_O (i|s) =$ occurrence period SI index, representing the ratio of dollar values at occurrence period $u$ to those at occurrence time 0
    \item[$\centerdot$] $n_i =$ number of claims occurring in occurrence period $i$
    \item[$\centerdot$] $r=$ identification number of claims occurring in occurrence period $i\ (r= 1,2, \dots ,N_i)$
    \item[$\centerdot$] $u_{ir}=$ occurrence time of claim $r$ of occurrence period $i$ (N.B. we have $i-1 < u_{ir} <i$)
    \item[$\centerdot$] $s_{ir}=$ size of claim $r$ of occurrence period $i$ without allowance for inflation
    \item[$\centerdot$] $v_{ir}=$ delay from occurrence to notification of claim $r$ of occurrence period $i$ (N.B. the notification time is $u_{ir}+v_{ir}$)
    \item[$\centerdot$] $w_{ir}=$ delay from notification to settlement of claim $r$ of occurrence period $i$ (N.B. the settlement time is $u_{ir}+v_{ir}+w_{ir}$)
    \item[$\centerdot$] $m_{ir}=$ number of partial payments in respect of claim $r$ of occurrence period $i$
    \item[$\centerdot$] $s_{ir}^{(m)}=$ size of the $m$-th partial payment in respect of claim r of occurrence period $i,m=1,2,\dots ,m_{ir}$
    \item[$\centerdot$] $p_{ir}^{(m)}=s_{ir}^{(m)}⁄\rev{s_{ir}}=$ proportion of claim amount $s_{ir}$ paid in the $m$-th partial payment
    \item[$\centerdot$] $d_{ir}^{(m)}=$ the inter-partial delay between from the epoch of the $(m-1)$-th to the $m$-th partial payment of claim $r$ of occurrence period $i$, with the convention that $d_{ir}^{(0)}=0$, corresponding to notification date (by convention, the $0$-th ``payment” is in fact the notification, without actual payment)
    \item[$\centerdot$] $\bar{t}_{ir}^{(m)}= u_{ir} + v_{ir} + d_{ir}^{(1)} + \dots + d_{ir}^{(m)}=$ the epoch of the $m$-th partial payment
\end{arcitem}

All of these quantities from $n_i$ onward, but except $r$, are realizations of random variables. The random variables themselves are denoted in the same way but with the primary symbol in upper case. For example, $S_{ir}$ denotes the random variable whose realization is $s_{ir}$.

\subsection{Incurred losses}
\noindent The following additional notation, specific to case estimates, is introduced:
\begin{arcitem}
    \item[$\centerdot$] $\tau=$ a generic variate denoting (continuous) time elapsed from claim notification
    \item[$\centerdot$] $w_{ir}^{(m)}=d_{ir}^{(1)} + \dots + d_{ir}^{(m)} =$ delay from notification to epoch of $m$-th partial payment $(m = 1, 2, \dots , m_{ir} )$ in the case of claim \rev{$r$} of occurrence period $i$
    \item[$\centerdot$] $w_{ir}=w_{ir}^{(m_{ir})}$ delay from notification to settlement in the case of claim $r$ of occurrence period $i$
    \item[$\centerdot$] $w_{ir}^{(m_{ir}-1)}=$ delay from notification to epoch of the penultimate partial payment (i.e. the final major payment) in the case of claim $r$ of occurrence period $i$
    \item[$\centerdot$] $m_{ir\tau } =$ largest integer $m$ for which $d_{ir}^{(1)} + \dots + d_{ir}^{(m)} \leq \tau$
    \item[$\centerdot$] $c_{ir}(\tau) = \sum_{m=1}^{m_{ir\tau }} s_{ir}^{(m)}=$ cumulative claim payments up to and including delay $\tau$ from notification in respect of claim $r$ of occurrence period $i$
    \item[$\centerdot$] $y_{ir}(\tau)=$ case estimate of ultimate incurred loss at delay $\tau$ from notification in respect of claim $r$ of occurrence period $i$
    \item[$\centerdot$] $x_{ir}(\tau) = y_{ir}(\tau ) - c_{ir} (\tau)=$ case estimate of outstanding claim payments at delay $\tau$ from notification in respect of claim $r$ of occurrence period $i$
    \item[$\centerdot$] $k_{ir}^{\text{Ma}}=$ number of major revisions of incurred loss during the life of claim $r$ of occurrence period $i$
    \item[$\centerdot$] $k_{ir}^{\text{Mi}}=$ number of minor revisions of incurred loss during the life of claim $r$ of occurrence period $i$
    \item[$\centerdot$] $\tau_{irl}^{\text{Ma}}=$ delay from notification to the epoch of $l$-th major revision of incurred loss during the life of claim $r$ of occurrence period $i$
    \item[$\centerdot$] $\tau_{irl}^{\text{Mi}}=$ delay from notification to the epoch of $l$-th minor revision of incurred loss during the life of claim $r$ of occurrence period $i$
    \item[$\centerdot$] $g_{irl}^{\text{Ma}}=$ revision multiplier at the $l$-th major revision of \textbf{incurred loss} during the life of claim $r$ of occurrence period $i$, causing $y_{ir} (\rev{\tau^{-}})$ to be replaced by $y_{ir}(\tau )= g_{irl}^{\text{Ma}} y_{ir} (\rev{\tau^{-}})$ at $\tau = \tau_{irl}^{\text{Ma}}$\rev{, where $\tau^{-}$ denotes $\lim_{\epsilon \downarrow 0}(\tau-\epsilon)$}
    \item[$\centerdot$] $g_{irl}^{\text{Mi}}=$ revision multiplier at the $l$-th minor revision of \textbf{outstanding claim payments} during the life of claim $r$ of occurrence period $i$, causing $x_{ir} (\rev{\tau^{-}})$ to be replaced by $x_{ir} (\tau)=g_{irl}^{\text{Mi}} x_{ir} (\rev{\tau^{-}})$ at $\tau = \tau_{irl}^{\text{Mi}}$
\end{arcitem}

\noindent Finally, note:
\begin{arcitem}
\item Revisions are assumed to occur at precisely the epoch $\tau$, i.e. the revision has not occurred at $\rev{\tau^{-}}$.
\item According to the explanation in Section~\ref{ssec:case estimates descript}, a major revision applies a factor to estimated incurred loss, whereas a minor revision applies a factor to estimated outstanding loss.  
\end{arcitem}

\section{Architecture of the claims process}\label{sec:architecture}

\subsection{Modular structure}\label{ssec:architecture claim process}

\noindent The claim process for claim $r$ of occurrence period $i$ is envisaged as consisting of the following modules:
\begin{description}
\item[Module 1:] Claim occurrence date; \label{module: module1}
\item[Module 2:] Claim size without allowance for inflation;\label{module: module2}
\item[Module 3:] Claim notification date;\label{module: module3}
\item[Module 4:] Claim settlement date;\label{module: module4}
\item[Module 5:] Number of partial payments;\label{module: module5}
\item[Module 6:] Sizes of partial payments without allowance for inflation;\label{module: module6}
\item[Module 7:] Distribution of payments over time;\label{module: module7}
\item[Module 8:] Claim inflation;\label{module: module8}
\item[Module 9:] Major revisions of incurred losses (number of revisions, distribution of revisions over time, and sizes of revisions);\label{module: module9}
\item[Module 10:] Minor revisions of incurred losses (number of revisions, distribution of revisions over time, and sizes of revisions);\label{module: module10}
\item[Module 11:] Development of case estimates, with the option to include inflation. \label{module: module11}
\end{description}


Modules 1 to 8 are present in the original version of \texttt{SynthETIC} \citep*{AvTaWaWo21,R-SynthETIC} and were described there.  Modules 9 to 11 are additional, and relate specifically to the simulation of case estimates in \texttt{SPLICE}. The present section details them.

\citet*{R-SynthETIC} commented on the modular structure of \texttt{SynthETIC}. While the algebraic structure of \texttt{SynthETIC} has always been quite general, the authors noted that there could be cases where change of functional dependencies would be required. The modularity of \texttt{SynthETIC} has ensured that the user could unplug any one and replace with a version modified to his/her own purpose.

The modular structure of \texttt{SynthETIC} is retained in \texttt{SPLICE}. Each type of revision of incurred losses (major or minor) are simulated in three sub-modules: frequency of revisions, their distribution in time, and sizes of revision factors; see Sections \ref{S_M9} to \ref{S_M10} and the subsections therein. The functional structure is designed to be general, and many users should be able to adopt it with changes to parameters but not algebraic structure.  However, in cases in which some change of structure is necessary, the modules can be unplugged and replaced with ease.

The sequence of Module 1 to Module 11 must be preserved, because each module typically relies on the output of prior modules.  Examples of such dependencies are provided in Section \ref{sssec: intramodel dependencies} \citep*[and in Section 4.3 of][]{AvTaWaWo21}.

\subsection{Module 9: Major revisions} \label{S_M9}
This section introduces a suite of functions that work together to simulate, in sequential order, (1) number of major revisions of incurred loss (\texttt{claim\_majRev\_freq}), (2) distribution of major revisions over time (\texttt{claim\_majRev\_time}), and (3) factors of major revisions (\texttt{claim\_majRev\_size}), for each of the claims occurring in each of the occurrence periods.

In particular, \texttt{claim\_majRev\_freq()} sets up the structure of the output for major revisions: a nested list such that the $j$th component of the $i$th sub-list is a list of information on major revisions of the $j$th claim of occurrence period $i$. The ``unit list" (i.e. the smallest, innermost sub-list that is unique to each individual claim) consists of the components in Table~\ref{tab:majRev}.

\begin{table}[!htbp]
\centering
\begin{tabular}{p{3cm}p{12cm}}
\toprule
Name & Description \\ \midrule
\verb=majRev_freq= & Number of major revisions of incurred loss; see \ref{ssec:majRev_freq}. \\ \midrule
\verb=majRev_time= & Epochs of major revisions (time measured from claim notification); see  \ref{ssec:majRev_time}. \\ \midrule
\verb=majRev_factor= & Major revision multiplier of \textbf{incurred loss}; see \ref{ssec:majRev_size}. \\ \midrule
\verb=majRev_atP= & An indicator, 1 if the last major revision occurs at the time of the last major payment (i.e. second last payment), 0 otherwise; see \ref{ssec:majRev_time}. \\ \bottomrule
\end{tabular}
\caption{List structure of major revisions of incurred loss}
\label{tab:majRev}
\end{table}

\subsubsection{Number of major revisions of incurred loss \emph{(\texttt{claim\_majRev\_freq})}} \label{ssec:majRev_freq}

The number of major revisions $k_{ir}^{\text{Ma}}$ is the realization of a random variable $K_{ir}^{\text{Ma}}$ with df $F_{K|s}^{\text{Ma}}(k;s)$, specified on input as a function of $k$, and possibly dependent on claim size $s$.

The default version of $F_{K|s}^{\text{Ma}}(k;s)$ is set out in Appendix~\ref{appendix:A}. In addition to the default parametrization, \texttt{SPLICE} supports a full range of user-specified alternative sampling distributions, with modifiable dependencies on other claim variates, as is the case for all modules that follow.

Users of the package can choose any suitable sampling distribution through the arguments \texttt{rfun} (a random sampling function) and \texttt{paramfun} (parameters for the random sampling function) to better serve their own testing purposes. As illustrated in Figure~\ref{fig:rfun} below, \texttt{rfun} defines the functional form of the distribution and can be any of pre-defined distributions in base R, or more advanced ones from other packages such as \texttt{actuar} \citep{R-actuar}, or any proper user-defined function, while \texttt{paramfun} creates the link between the previously simulated quantities and the parameters of \texttt{rfun}. Here, ``parameters" should be interpreted in a very general sense: for a specific parametric distribution---e.g., Weibull, this can simply be the shape and scale parameters of the distribution, defined as a function of already simulated quantities---e.g. \texttt{claim\_size} (from \texttt{SynthETIC}). For a user-defined \texttt{rfun} taking any other arguments, \texttt{paramfun} should output the required \texttt{rfun} arguments as a function of the already simulated quantities.

\begin{figure}[!htbp]
    \centering
    \includegraphics[width=0.9\textwidth]{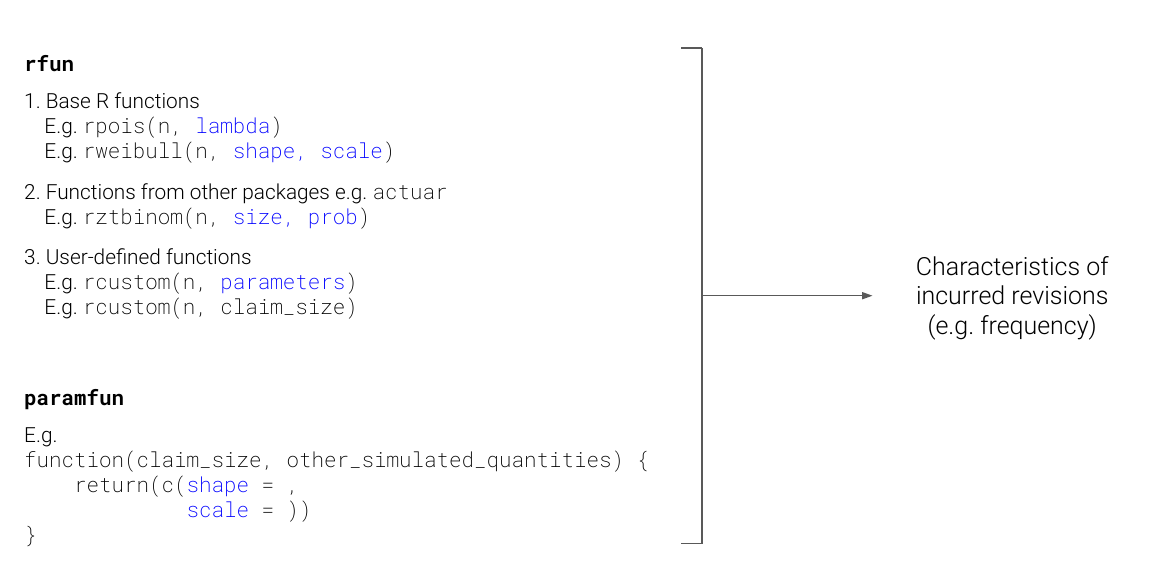}
        \caption{\texttt{SPLICE} implementation}
        \label{fig:rfun}
\end{figure}

The \texttt{SPLICE} vignette, which can be accessed via

\texttt{R> vignette("SPLICE-demo", package = "SPLICE")}

or online at \url{https://CRAN.R-project.org/package=SPLICE} \citep*{R-SPLICE}, includes illustrative examples on using a zero-truncated Poisson distribution with both default and modified dependence structures \rev{(e.g. adding the dependence of $K_{ir}^{\text{Ma}}$ on the number of partial payments $m_{ir}$ of the claim in addition to its default dependence on claim size $s_{ir}$)}.

\subsubsection{Distribution of major revisions over time \emph{(\texttt{claim\_majRev\_time})}} \label{ssec:majRev_time}
As noted in Principle~\ref{principles: principle6}, each claim experiences a major revision at notification.  The following remarks relate to any subsequent major revisions.  These occur only for claims with $m_{ir} \geq 4$.

If $l=k_{ir}^{\text{Ma}}$, that is, when considering the last major revision $g_{irl}^{\text{Ma}}$, the probability that this coincides with the penultimate claim payment (i.e. the final major payment) is $P \left( \tau_{irl}^{\text{Ma}}=w_{ir}^{(m_{ir}-1)} |s_{ir} \right)$, possibly dependent on claim size.  In this event (\texttt{majRev\_atP == 1}), the $\tau_{irl}^{\text{Ma}},l=2, \dots ,k_{ir}^{\text{Ma}}-1$ are realizations of a random variable $T^{\text{Ma}1}$ with df $F_{T|w}^{\text{Ma}1} (\tau ;w)$ specified on input as a function of $\tau$.

In the event that the last major revision does not coincide with the penultimate claim payment (\texttt{majRev\_atP == 0}), the $\tau_{irl}^{\text{Ma}},l=2, \dots , k_{ir}^{\text{Ma}}$ are realizations of a random variable $T^{\text{Ma}2}$ with df $F_{T|w}^{\text{Ma}2} (\tau;w)$ specified on input as a function of $\tau$.

\subsubsection{Factors of major revisions of incurred loss (\emph{\texttt{claim\_majRev\_size})}} \label{ssec:majRev_size}
As noted in Principle~\ref{principles: principle6}, each claim experiences a major revision at notification.  The following remarks relate to any subsequent major revisions.

The factor by which estimated incurred loss is adjusted at a major revision $g_{irl}^{\text{Ma}}$ is the realization of a random variable $G_{irl}^{\text{Ma}}$ with df $F_{G|h}^{\text{Ma}} (g;h)$, specified on input as a function of $g$, and possibly dependent on the history of major revisions $h_l$ preceding the $l$-th.

The default version of $F_{G|h}^{\text{Ma}} (g;h)$ is set out in Appendix~\ref{appendix:A}. The \texttt{SPLICE} vignette illustrates how to deploy alternative sampling assumptions.

\subsection{Module 10: Minor revisions} \label{S_M10}
For the simulation of minor revisions, we treat separately the case of revisions that occur simultaneously with a partial payment and the ones that do not (Principle 4 in Section~\ref{ssec:case estimates descript}).

Similar to the case of major revisions, the suite of functions under this heading run in sequential order to simulate (1) number of minor revisions of incurred loss (\texttt{claim\_minRev\_freq}), (2) distribution of minor revisions over time (\texttt{claim\_minRev\_time}), and (3) factors of minor revisions (\texttt{claim\_minRev\_size}), for each of the claims occurring in each of the occurrence periods.

Analogous to major revisions, \texttt{claim\_minRev\_freq()} sets up the structure of the output minor revisions: a nested list such that the $j$th component of the $i$th sub-list is a list of information on minor revisions of the $j$th claim of occurrence period $i$. The ``unit list" consists of the components in Table~\ref{tab:minRev}.

\begin{table}[!htbp]
\centering
\begin{tabular}{p{4cm}p{11cm}}
\toprule
Name & Description \\ \midrule
\verb=minRev_atP= & A logical vector indicating whether there is a minor revision at each partial payment; see \ref{ssec:minRev_freq}. \\ \midrule
\verb=minRev_freq_atP= (\verb=minRev_freq_notatP=) & Number of minor revisions that occur (or do not occur) simultaneously with a partial payment. \texttt{minRev\_freq\_atP} is numerically equal to the sum of \texttt{minRev\_atP}; see \ref{ssec:minRev_freq}. \\ \midrule
\verb=minRev_time_atP= (\verb=minRev_time_notatP=) & Epochs of minor revisions that occur (or do not occur) simultaneously with a partial payment (time measured from claim notification); see~\ref{ssec:minRev_time}. \\ \midrule
\verb=minRev_factor_atP= (\verb=minRev_factor_notatP=) & Minor revision multiplier of \textbf{outstanding claim payments} for revisions at partial payments and at any other times, respectively; see~\ref{ssec:minRev_size}. \\ \bottomrule
\end{tabular}
\caption{List structure of minor revisions of incurred loss}
\label{tab:minRev}
\end{table}

\subsubsection{Number of minor revisions of incurred loss \emph{(\texttt{claim\_minRev\_freq})}} \label{ssec:minRev_freq}
The number of minor revisions $k_{ir}^{\text{Mi}}$ consists of two components: $k_{ir}^{\text{Mi}}=k_{ir}^{\text{Mi}1}+k_{ir}^{\text{Mi}2}$, where those counted in $k_{ir}^{\text{Mi}1}$ are simultaneous with a partial payment (possibly final payment), and those counted in $k_{ir}^{\text{Mi}2}$ are not. 

The variate $k_{ir}^{\text{Mi}1}=\sum_{l=1}^{M_{ir}} b_l$, where the $b_l$ are realizations of independent Bernoulli variates $B_l$, each corresponding to the $l$-th partial payment, and having df $F_{B|l}(b)$.

The variate $k_{ir}^{\text{Mi}2}$ is the realization of a random variable $K_{ir}^{\text{Mi}2}$ with df $F_{K|w}^{\text{Mi}2}(k;w)$, specified on input as a function of $k$, and possibly dependent on settlement delay $w$.

The default versions of $F_{B|l} (b)$ and $F_{K|w}^{\text{Mi}2} (k;w)$  are set out in Appendix~\ref{appendix:A}.

\subsubsection{Distribution of minor revisions over time \emph{(\texttt{claim\_minRev\_time})}} \label{ssec:minRev_time}
If a minor revision occurs in conjunction with a partial payment, its epoch is equal to the epoch of the payment.

If a minor revision occurs at an epoch other than those of partial payments, then the $\tau_{irl}^{\text{Mi}}$ are realizations of a random variable $T^{\text{Mi}}$ with df $F_{T|w}^{\text{Mi}} (\tau;w)$ specified on input as a function of $\tau$.

Major and minor revisions cannot occur simultaneously. In the event that they are simulated to do so (which will only ever occur at the last major payment), the major revision takes precedence, and the minor revision is discarded. This adjustment is made at the consolidation step (i.e. Module 11).

\subsubsection{Factors of minor revisions of incurred loss \emph{(\texttt{claim\_minRev\_size})}} \label{ssec:minRev_size}
The factor by which case estimate is adjusted at a minor revision $g_{irl}^{\text{Mi}}$ is the realization of a random variable $G_{irl}^{\text{Mi}}$ with df $F_{G|w,\tau}^{\text{Mi}} (g;w,\tau)$, specified on input as a function of $g$, and possibly dependent on the delay $w$ from notification to settlement and the delay $\tau$ from notification to the subject minor revision.

The default version of $F_{G|w,\tau}^{\text{Mi}} (g;w,\tau)$ is set out in Appendix~\ref{appendix:A}. Again we refer to the \texttt{SPLICE} vignette and package documentation for illustrations of alternative parametrizations \citep*{R-SPLICE}.

\subsection{Module 11: Computation of case estimates \emph{(\texttt{claim\_history})}} \label{S_11}

\subsubsection{Without inflation} \label{S_11woinfl}

Initially, base inflation will be ignored.  All case estimates will be computed in values corresponding to time $\bar{t} = 0$, i.e. the commencement of the first occurrence period.

For each claim, claim size (before base inflation) is simulated within the Payments section of \texttt{SynthETIC}.  By Principle~\ref{principles: principle11}, the case estimate at settlement (again before base inflation) must coincide with it.  Symbolically,
\begin{equation}\label{eqn: sir}
    y_{ir}(w_{ir}) = s_{ir}    .
\end{equation}
In order to ensure this identity, it is necessary to simulate case estimates in reverse chronological time.  One commences by setting $y_{ir} (w_{ir} )$ in accordance with \ref{eqn: sir}, then calculating  $y_{ir} (w_{ir}-0)$.  This will be equal to $y_{ir} (w_{ir} )$ if no revision of incurred amount occurs at settlement. Otherwise, it will be calculated by means of (\ref{eqn: yir mi}) below.

Note that $y_{ir} (\tau)=y_{ir} (w_{ir}-0)$ for $\tau$ equal to the delay from notification to the epoch of the last revision (major or minor) strictly prior to settlement.  From this $y_{ir} (\rev{\tau^{-}})$ is calculated, with allowance for the revision by either (\ref{eqn: yir ma}) or (\ref{eqn: yir mi}).  Working in reverse order in this way, one calculates $y_{ir} (\tau)$ for all $0 \leq \tau \leq w_{ir}$.  The value of $y_{ir} (0)$ arrived at is the initial case estimate (at notification) for the claim.

The relations used to calculate a pre-revision case estimate from post-revision case estimate at epoch $\tau$ are \textbf{initially} as follows:
\begin{equation}\label{eqn: yir ma}
y_{ir}(\rev{\tau^{-}}) = \frac{y_{ir}(\tau)}{g_{irl}^{\text{Ma}}}
\end{equation}
if the $l$-th major revision occurs at epoch $\tau$; or
\begin{equation}\label{eqn: yir mi}
y_{ir}(\rev{\tau^{-}}) = c_{ir}(\rev{\tau^{-}}) + \frac{y_{ir}(\tau) - c_{ir}(\rev{\tau^{-}})}{g_{irl}^{\text{Mi}}}
\end{equation}
if the $l$-th minor revision occurs at epoch $\tau$.

When a minor revision coincides with a partial payment, there is a need to define whether the revision of outstanding claims occurs first and is then followed by the payment, or vice versa.  Note that (\ref{eqn: yir mi}) is equivalent to
\begin{equation}\label{eqn: yir mi 2}
    y_{ir}(\tau) = c_{ir}(\rev{\tau^{-}}) + g_{irl}^{\text{Mi}} x_{ir}(\rev{\tau^{-}}),
\end{equation}
which means that the revision occurs first.

\begin{figure}[!htbp]
    \centering
    \includegraphics[width=\textwidth]{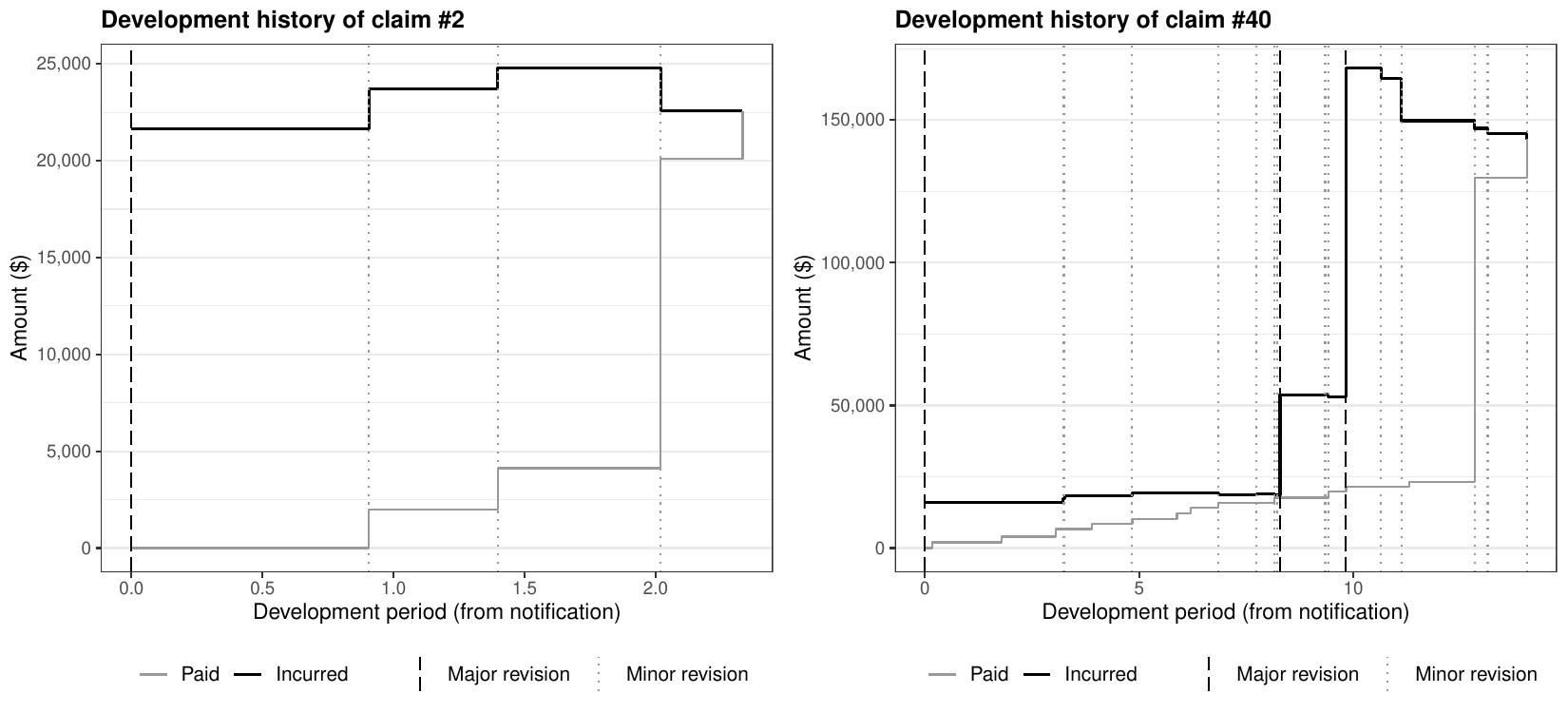}
        \caption{Graphical representation of two sample claims from the \hyperref[ssec:example-implementation]{example implementation}}
        \label{fig:example claim history}
\end{figure}

As an illustrative example, Figure~\ref{fig:example claim history} visualizes the development of two sample claims (without inflation). The grey paths in the plots are simulated by \texttt{SynthETIC} \citep*{AvTaWaWo21}. They describe the full history of paid losses during the lifetime of a claim, which is taken as input by \texttt{SPLICE} for the generation of the incurred histories (black paths). For reasons explained above, \texttt{SPLICE} works backward from the settlement of the claim, sets the case estimate at settlement equal to the claim size, and only updates the case estimates when a revision of incurred loss is projected to occur. The jumps in the black paths thus correspond to the points of major or minor revisions. We remark that some of the jumps may coincide with a partial payment (e.g. all three partial payments for claim \#2 result in a simultaneous minor revision; left panel in Figure~\ref{fig:example claim history}). The computation of case estimates in such cases is governed by Equation~\eqref{eqn: yir mi 2}, and detailed in Appendix \ref{appendix:data excerpt}. For details of the two claim records, we refer to the data excerpt in Appendix~\ref{appendix:data excerpt}.

It is stated above that \eqref{eqn: yir ma}--\eqref{eqn: yir mi 2} apply only ``initially”, because adjustments may be required to deal with \textbf{constrained cases}. For example, a large value of $g_{irl}^{\text{Ma}}$ in (\ref{eqn: yir ma}) could force $y_{ir} (\rev{\tau^{-}})$ below $c_{ir} (\rev{\tau^{-}})$, in which case $x_{ir} (\rev{\tau^{-}})$ would be negative.

By convention, case estimates should be strictly positive.  This is enforced a \textit{fortiori} by requiring that 
\begin{equation}\label{eqn: yir constraint}
    \kappa y_{ir}(\rev{\tau^{-}}) \geq c_{ir}(\rev{\tau^{-}}),
\end{equation}
for all $i$, $r$ and $\tau < \tau_{ir}$, where $0 < \kappa < 1$ is a constant. 
Thus, if (\ref{eqn: yir ma}) or (\ref{eqn: yir mi}) yields a value of $y_{ir} (\rev{\tau^{-}})$ that breaches (\ref{eqn: yir constraint}), then it is corrected to 
\begin{equation}\label{eqn: yir constraint 2}
    y_{ir}(\rev{\tau^{-}}) = \kappa^{-1} c_{ir}(\rev{\tau^{-}}).
\end{equation}
Constraint (\ref{eqn: yir constraint}) has an equivalent form, that is useful in application to (\ref{eqn: yir mi}).  This is
\begin{align*}
    c_{ir} (\rev{\tau^{-}}) \leq \kappa y_{ir} (\rev{\tau^{-}}) &= \kappa(c_{ir} (\rev{\tau^{-}}) + x_{ir} (\rev{\tau^{-}})) \\
    &= \kappa \left( c_{ir} (\rev{\tau^{-}}) + \frac{x_{ir} (\tau)}{g_{irl}^{\text{Mi}}} \right),
\end{align*}
from which follows
\begin{equation}\label{E_47}
    g_{irl}^{\text{Mi}} \leq \frac{\kappa}{1-\kappa}\frac{x_{ir}(\tau)}{c_{ir}(\rev{\tau^{-}})}.
\end{equation}

\subsubsection{Base inflation adjustment}
The allowance for inflation in case estimates is explained in Section~\ref{ssec: treatment of inflation}. At each revision of incurred loss, an adjustment is made for the base inflation that has occurred since the previous revision.  Within the simulator these adjustments are made in reverse chronological order.

Consider an adjustment for the period between a revision at delay $\tau^*$ and $\rev{\tau^{-}} > \tau^*$, where it is possible that $\tau^*=0$, and also possible that $\tau=\tau_{ir}$.  Let delay $\tau$ correspond to time $\bar{t}$, i.e. $\bar{t}=u_{ir}+v_{ir}+\tau$.  Similarly, let delay $\tau^*$ correspond to time $\bar{t}^* = \bar{t} - (\tau - \tau^*)$.  Then the adjustment factor (in reverse time) is $f(\bar{t}^*)⁄f(\bar{t})$.

As an example, if $y_{ir} (\rev{\tau^{-}})$ is computed by either (\ref{eqn: yir ma}) or (\ref{eqn: yir mi}), and the revision immediately prior to delay $\tau$ is a major revision at $\tau^*$, then (\ref{eqn: yir ma}) at that epoch is replaced by
\begin{equation}
    y_{ir}(\tau^*) = y_{ir}(\rev{\tau^{-}}) \frac{f(\bar{t}^*)}{f(\bar{t})} = \frac{y_{ir}(\tau)}{g_{irl}^{\text{Ma}}}\frac{f(\bar{t}^*)}{f(\bar{t})}.
\end{equation}

\subsection{\rev{The treatment of ``manual adjustments''}}

\rev{``Manual adjustments'' occur twice within the default version of \texttt{SPLICE}, firstly in the requirement that minor and major revisions cannot occur simultaneously (Section~\ref{ssec:minRev_time}), and secondly in the requirement of strict positivity of case estimates (Section~\ref{S_11woinfl}).}

\rev{It may happen that a user has cause to fit the simulation model to a real data set by maximum likelihood or by optimization of some other statistical criterion. In this case, a pure mathematical statement of the model will be required, and any ``manual adjustments'' will require accommodation within it.}

\rev{It should first be pointed out that the default model is quite optional, and the user is free to modify it in any form desired, eliminating ``manual adjustments'' if necessary. However, it can also be pointed out that the default model is in fact convertible to pure mathematical form without manual adjustments. }

\rev{Consider, for example, the requirement that major and minor revisions not coincide. Coincidence could occur in the default implementation only if a minor and major revisions were simulated to coincide with the last major payment, where $\tau=\tau_{irl}^{\text{Ma}}=\omega_{ir}^{(m_{ir}-1)}$} (see Section~\ref{ssec:minRev_time}).
The df of the minor revision factor $g_{irl}^{\text{Mi}}$, according to Section~\ref{ssec:minRev_size}, is $F_{G|\omega,\tau}^{\text{Mi}}$, in other words dependent on the epoch of the minor revision in question. The over-riding of the minor revision could be incorporated into this df by extension of the conditioning of $F_{G|\omega,\tau}^{\text{Mi}}$ to $F_{G|\omega,\tau,{\tau_{irl}^{\text{Ma}}},{\omega_{ir}^{(m_{ir}-1)}}}^{\text{Mi}}$ and stipulation within $F_{G|\omega,\tau,{\tau_{irl}^{\text{Ma}}},{\omega_{ir}^{(m_{ir}-1)}}}^{\text{Mi}}$ that $g_{irl}^{\text{Mi}}=1$ if $\tau_{irl}^{\text{Ma}}=\omega_{ir}^{(m_{ir}-1)}$.

\rev{This is an extremely ugly expression of the model, and is not recommended for model description. It does illustrate, however, the a way in which the over-riding of a minor revision by a major one can be expressed in the form of a genuine statistical model.}

\rev{A similar device can be used to enforce the strict positivity of case estimates, as is achieved by \eqref{eqn: yir constraint}--\eqref{E_47}. Again the device consists of expanding the conditionality of the random variable $G_{irl}^{\text{Mi}}$.}

\subsection{Aggregation of output}

Consistent with \texttt{SynthETIC}, \texttt{SPLICE} provides both individual claim and aggregate output.  In the latter, transactions are aggregated by accident and development period, where the duration of the periods may be chosen to any desired level of granularity (e.g. users who choose to work with calendar months can aggregate the transactions by month, quarter, or year). \texttt{SPLICE} by default uses accident and development quarters. The aggregate of case estimates for any particular development quarter includes the case estimates of all relevant individual claims at the \textbf{end of the quarter}. If a claim’s incurred loss is revised more than once during the quarter, only the estimated incurred loss after the last of those revisions will be reflected in the aggregate. Likewise, if the user chooses to summarize the claims on a yearly level, then the claim triangles will only capture the latest estimated incurred loss at the end of each year. As an illustration, Appendix~\ref{appendix:triangles} shows the cumulative incurred loss triangle of the example implementation described in Section~\ref{ssec:example-implementation}.  

\subsection{Out-of-bounds transactions}

In this sub-section, a ``transaction" includes occurrence, notification, settlement, a payment, or a case estimate revision.  

Sometimes, simulated transactions will take place beyond the end of the last development period. 
This out-of-bounds issue can potentially occur with all types of transaction \citep*[see also][]{AvTaWaWo21}. \texttt{SPLICE} treats those cases according to the following convention.

The bounds on development periods are ignored throughout that simulation, except in any aggregation of output by development period or addition of inflation, where any out-of-bounds transactions are counted as if at the end of the limiting development period $I$ (where development periods are numbers 1, 2, ...).

In the specific case of incurred losses, the simulated epoch of occurrence of any incurred loss revision is maintained throughout the simulation of details of the claim concerned, other than in the exceptions noted below. For example, if a minor revision occurs at development time $j>I$, and sizes of the minor revision multipliers depend on the epochs of the subject revisions, then the simulated value of $j$ will be used in the simulation of those revision multipliers.  

The epoch of revision is varied only at the stage where case estimates are assigned to development periods for the purpose of either tabulation or addition of inflation.  In this context, the revision is assumed to have occurred at the end of development period $I$.  In short, the integrity of epochs of transaction, and of any dependency on these epochs, is maintained throughout, with the sole exceptions of aggregation of out-of-bounds settlements and adjustment for inflation.

\subsection{Data features} \label{S_datafeatures}

\texttt{SPLICE} inherits the claim payment structure from \texttt{SynthETIC}, which has been structured to resemble a real Auto Liability portfolio (``reference portfolio"). We refer to Section 4.3 of \citet*{AvTaWaWo21} for a review of the data features of the portfolio related to claim payments; see also \citet*{TaMcGr03,TaMc04,McG07b,McTaMi18}.

As a result of the features alluded to in the previous paragraph, the portfolio behaviour could change over time with respect to claim payments \citep*[see][for details]{AvTaWaWo21}. In contrast, the reference portfolio is not subject to time-heterogeneity in the behaviour of its incurred loss estimates.  There are, however, a few data features worthy of note.  Some are briefly described in Principles~\ref{principles: principle1}--\ref{principles: principle11}.  Broad details of one or two others are given immediately below.  Full detail appears in Appendix~\ref{appendix:A}.

\begin{principle}
The likelihood of a major revision at settlement increases with increasing claim size.\label{principles: principle12}
\end{principle}
\begin{principle}The timing of major revisions, other than those at settlement, is biased towards the early part of the claim’s lifetime.\label{principles: principle13}
\end{principle}
\begin{principle}The timing of minor revisions, other than those coincident with partial claim payments, is similarly biased.\label{principles: principle14}
\end{principle}

Although the generator of incurred loss estimates is time-homogeneous, as noted just above, it does not follow that the behaviour of those estimates will be without complexity.  As can be seen in Appendix~\ref{appendix:A}, the behaviour of the incurred loss estimates is dependent on that of the claim payments.  The latter are time-heterogeneous, and some of the consequent complexity can be transmitted to the incurred loss estimates (see Sections~\ref{ssec: CL comparison} and \ref{sssec: intramodel dependencies}).

\section{\rev{Application of \texttt{SPLICE}}}\label{sec:SPLICE-application}
\subsection{\rev{Example implementation of \texttt{SPLICE} with default parametrisation}}\label{ssec:example-implementation}
\subsubsection{\rev{Modular implementation in \texttt{R}}} \label{sssec:example-code}
\citet*{AvTaWaWo21} performed an example simulation of claim payments in accordance with a detailed specification given there, with principal features of the experience similar to those of the reference portfolio.  The generated experience covered 40 occurrence quarters, each tracked for 40 development quarters, with detailed transactional records.

\texttt{SPLICE}  has been used to extend that simulation to include case estimates.  Simulated paid losses remain unchanged from the earlier paper. The transactional simulation output now comprises key dates, and both claim payments and revisions of estimated incurred losses.  A detailed specification of the modules involved in the simulation of case estimates is given in Appendix~\ref{appendix:A}. 

Below we present the code to generate the example data set that is included as part of the package and described in the following sections (\ref{ssec: CL comparison} and \ref{sssec: intramodel dependencies}). We refer to the \texttt{SPLICE} vignette and package documentation for details of the function usage \citep*{R-SPLICE}.

\begin{lstlisting}[language=R]
library(SPLICE)
set.seed(20201006)
test_claims <- SynthETIC::test_claims_object

# major revisions
major <- claim_majRev_freq(test_claims)
major <- claim_majRev_time(test_claims, major)
major <- claim_majRev_size(major)

# minor revisions
minor <- claim_minRev_freq(test_claims)
minor <- claim_minRev_time(test_claims, minor)
minor <- claim_minRev_size(test_claims, major, minor)

# development of case estimates
test <- claim_history(test_claims, major, minor)
test_inflated <- claim_history(
  test_claims, major, minor,
  base_inflation_vector = rep(1.02^(1/4) - 1, times = 80))

# transactional data
test_incurred_dataset_noInf <- generate_incurred_dataset(test_claims, test)
test_incurred_dataset_inflated <- generate_incurred_dataset(
  test_claims, test_inflated)
\end{lstlisting}

An excerpt of the transactional data set, \texttt{test\_incurred\_dataset\_noInf} generated from the above code, is included in Appendix~\ref{appendix:data excerpt}. Those results can easily be aggregated into triangles; see Appendix \ref{appendix:triangles}. 

Note that using the default set of parametrization (which has been loosely calibrated to the reference portfolio) does not require the user to input further arguments. In cases where alternative sampling distributions or dependence structures are desired, they can be easily incorporated using the \texttt{SPLICE} framework described in Figure~\ref{fig:rfun}.

\rev{The above is implementing the modules sequentially in a transparent way, but the dataset can  also be directly generated with a single function \texttt{generate\_data()}, for varying levels of complexity; see Section \ref{ssec:alternative-datasets}.}

\subsubsection{Comparison with chain ladder}\label{ssec: CL comparison}
The paid loss data simulated by \texttt{SynthETIC} was in substantial breach of the chain ladder assumptions \citep*{AvTaWaWo21}.  Section 6.2.1 of that paper demonstrated that those breaches translated into equally dramatic errors in the chain ladder’s forecast of a loss reserve.

The destructive data features of the earlier paper related essentially to heterogeneity of the claim payment simulation model over time. For example, large (resp. small) claims were affected by small (resp. large) rates of payment period SI.  This caused the observed profile of paid losses by development period to shift as one moved from one occurrence period to another.

In the present case, the incurred loss simulation model contains no such time-heterogeneities, as can be seen in its definition in Appendix~\ref{appendix:A}.  There is, nonetheless, the potential for some of the paid loss heterogeneities to induce incurred loss heterogeneities.  There are at least a couple of ways in which this could occur, but the dominant one is described as follows.

First, note that claim size and delay to settlement are positively associated \citep*{AvTaWaWo21}.  Then recall the remark just above on SI, as it affects large and small claims, and note that the effect of this would be to steadily shorten the distribution of delay to settlement with advancing occurrence period.  Next note that both major and minor revisions of incurred loss are triangularly distributed over a range of epochs determined by the delay to settlement.  It follows that the distribution of these epochs would steadily shorten with advancing occurrence period.

This has been investigated as follows:
\begin{enumerate}[label=(\alph*)]
    \item We aggregated the part of the simulated indvidual incurred data relating to quarters 1 to 40 into a $40 \times 40$ triangle.
    \item We derived a forecast of ultimate incurred loss (and hence outstanding claims up to development quarter 40) by applying a chain ladder. Note that there is little claim activity beyond development quarter 40.
    \item We compared this forecast with the ``actual” amount of outstanding claims, simulated for payment quarters 41 to 79.
\end{enumerate}

\begin{figure}[!htbp]
    \centering
    \includegraphics[width=.7\textwidth]{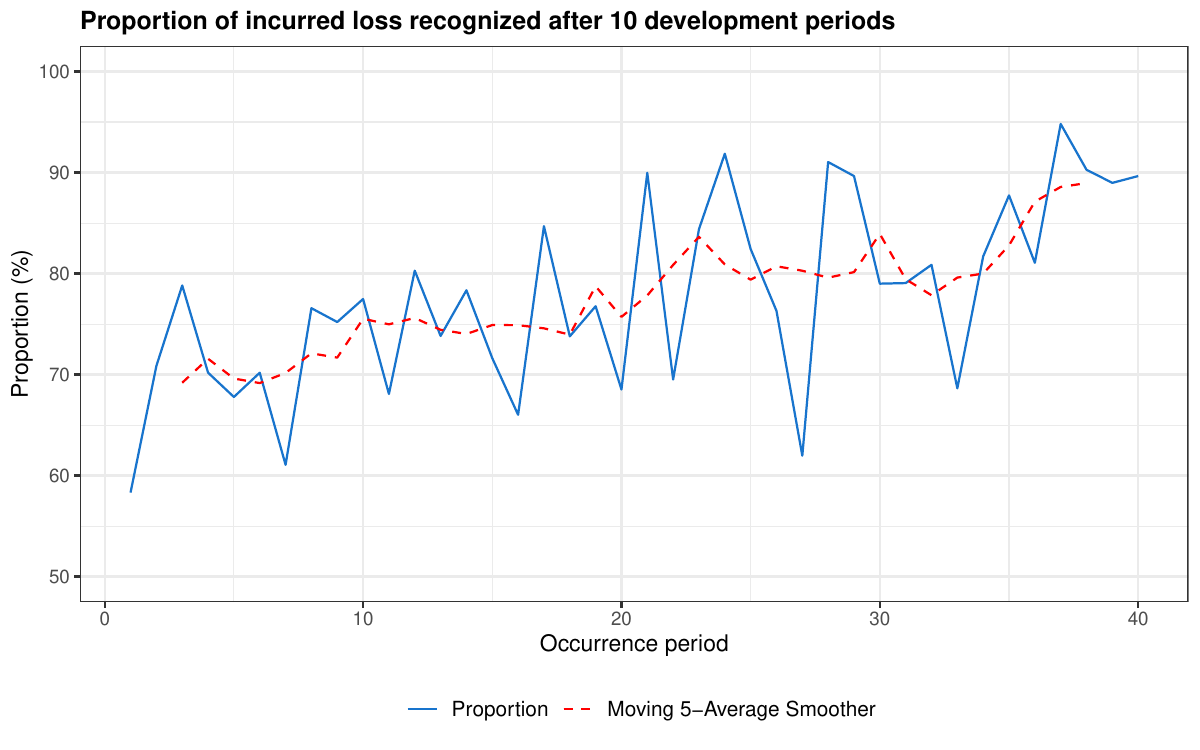}
        \caption{Time-heterogeneity of incurred losses over occurrence periods}
        \label{fig:incurred loss time-heterogeneity}
\end{figure}

The predicted shortening of the distribution of epochs of incurred loss revisions is indeed empirically confirmed by Figure~\ref{fig:incurred loss time-heterogeneity}, which plots as a solid line, for each of the 40 occurrence periods, the proportion of the ultimate incurred loss recognized in paid losses and case estimates at the end of the 10th development period.  The figure also shows as a dashed line the smoothed version of the plot with a moving 5-average as the smoother.  It can be seen that the proportion of incurred loss recognized increases from about 70\% to about 90\% over the span of the occurrence periods.

The forecast results are set out in Table~\ref{tab: synthetic forecast}, where the chain ladder exhibits persistent over-estimation of outstanding liability (see also Appendix~\ref{appendix:triangles}).  This is typical of the chain ladder in an environment of diminishing delay to recognition of incurred loss.  

\begin{table}[!htbp]
    \centering

    \begin{tabular}{P{2cm}P{3cm}P{3cm}P{3cm}}
         \toprule
         \multirow{2}{2cm}{\textbf{Occurrence Quarters}} & \multicolumn{3}{c}{\textbf{Estimated loss reserve}} \\
         \cmidrule{2-4}
         & \textbf{Target (Simulator)} & \textbf{Chain ladder estimate} & \textbf{Deviation of chain ladder from target} \\
         \midrule
         & \$M & \$M & \% \\
         & & & \\
         1 to 10 & 9.2 & 11.4 & 23 \\
         11 to 20 & 26.2 & 32.3 & 23 \\
         21 to 25 & 26.1 & 35.7 & 37 \\
         26 to 30 & 61.8 & 74.9 & 21 \\
         \midrule
         31 & 20.5 & 23.3 & 13 \\
         32 & 24.4 & 27.9 & 15 \\
         33 & 27.8 & 25.2 & -9 \\
         34 & 30.1 & 35.7 & 19 \\
         35 & 26.4 & 33.9 & 28 \\
         36 & 36.2 & 43.2 & 19 \\
         37 & 33.0 & 40.5 & 23 \\
         38 & 43.1 & 65.9 & 53 \\
         39 & 39.0 & 43.5 & 12 \\
         40 & 46.2 & 29.1 & -37 \\
         \midrule
         Total & 449.6 & 522.6 & 16 \\
         \bottomrule
    \end{tabular}
    \caption{Forecast claim costs based on synthetic data set}
    \label{tab: synthetic forecast}
\end{table}

\subsubsection{Intra-model dependencies}\label{sssec: intramodel dependencies}
Appendix~\ref{appendix:A} provides details of the general structure of incurred loss development in respect of a specific claim.  One point of note is that it does not follow a Markov process, i.e. a given claim transaction may depend on the entire history of the claim rather than on just its status at the point of the transaction in question.  For example, if a claim experiences a large upward revision, other than at notification, then any subsequent revision is unlikely to be large.

The reason for the inclusion of this condition in the simulator may be illustrated by means of an example.  Consider a claim that is initially estimated as of size \$50,000, and suppose that this estimate is subsequently revised to \$500,000, i.e. by a factor of 10.  It is possible that a further substantial revision would occur, to \$750,000 say, but a further revision by a factor 10 is unlikely.

\begin{figure}[!htbp]
    \centering
    \includegraphics[width=.7\textwidth]{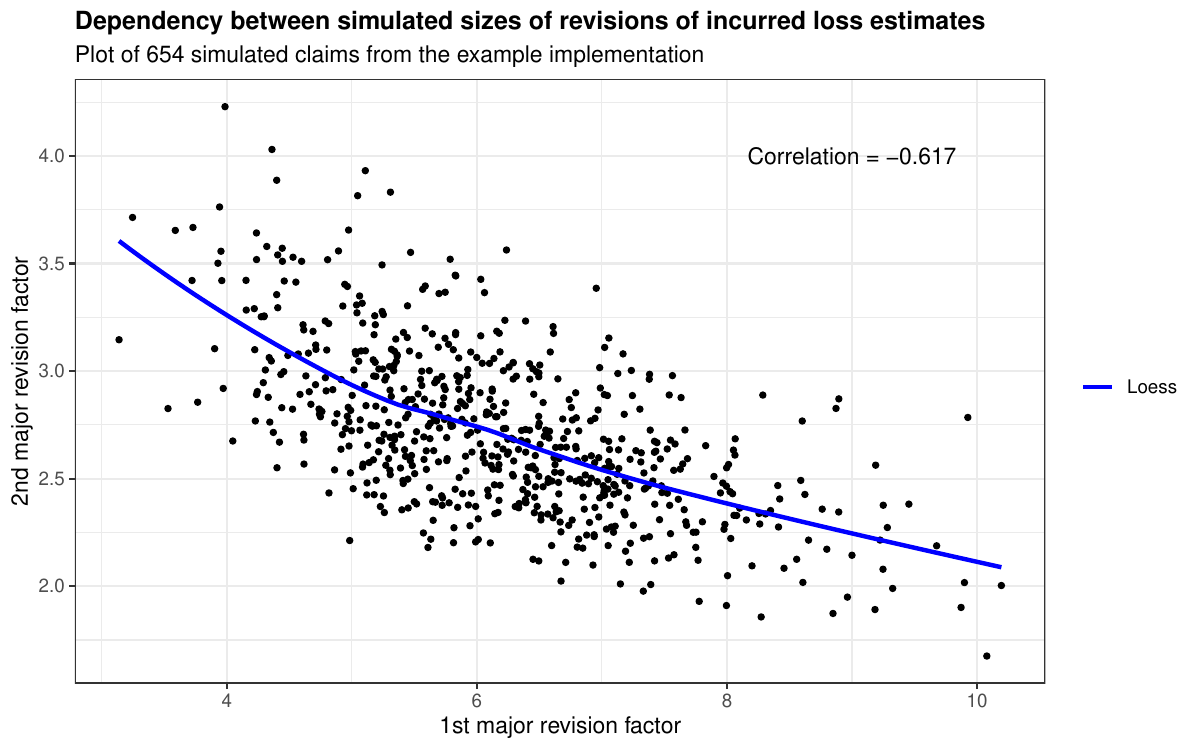}
        \caption{Dependency between simulated sizes of revisions of incurred loss estimates}
        \label{fig:revision size dependency}
\end{figure}

In fact, a negative association between the magnitudes of the first and second major revision factors (after notification) is evident from Appendix~\ref{appendix:A}.  This is illustrated by Figure~\ref{fig:revision size dependency}, which plots the major revision factors of 654 simulated claims from the data set described in Section~\ref{sssec:example-code} that experience two major revisions (in addition to the one at notification), with a superimposed loess curve. A ``revision factor”, as referred to here, is defined as the ratio by which the estimate of an individual claim’s incurred loss changes at any epoch.  Inflationary effects have been excluded. The empirical correlation of the two major revision factors is estimated to be $-0.617$ for the example implementation.  

\subsection{\rev{Alternative data sets of varying complexity}} \label{ssec:alternative-datasets}
Section~\ref{ssec: CL comparison} demonstrates the failure of the chain-ladder model to capture the time heterogeneities in the incurred losses of the default portfolio. However, the flexibility of \texttt{SPLICE} ensures that the user can generate data of almost any level of complexity. \rev{Indeed, as discussed in Section~\ref{ssec:background}, the user may generate a collection of data sets with varying levels of complexity and test the proposed model on the whole spectrum, in order to derive insights into its strengths and weaknesses.}

\begin{table}[htb]
\small
\centering
\caption{Descriptions of alternative data sets considered} \label{tab:scenarios}
\begin{tabular}{>{\raggedright\arraybackslash}p{2.5cm}|>{\raggedright\arraybackslash}p{2cm}>{\raggedright\arraybackslash}p{1.7cm}>{\raggedright\arraybackslash}p{2.3cm}>{\raggedright\arraybackslash}p{2cm}>{\raggedright\arraybackslash}p{3.5cm}}
\toprule
Complexity & 1 \hfill \hfill \linebreak (simplest) & 2 & 3 & 4 & 5 \hfill \hfill \linebreak (default, most complex) \\ \midrule
Notification delay dependence structure (Module 3) & Independent of all features & Depends on claim size & Depends on claim size & Depends on claim size & Depends on claim size \\ \midrule

Settlement delay dependence structure \hfill \hfill \linebreak (Module 4) & Depends on claim size & Depends on claim size & Depends on claim size and occurrence period (steady decline over time) & Depends on claim size & Depends on claim size and occurrence period (decline over first 20 periods, plus a structural break at time 20 for smaller claims) \\ \midrule

Base inflation \hfill \hfill \linebreak (Module 8, \hyperref[S_11]{11}) & Nil & Constant at 2\% p.a. & Constant at 2\% p.a. & Constant at 2\% p.a. & Constant at 2\% p.a. \\ \midrule

Calendar period superimposed inflation \hfill \hfill \linebreak (Module 8, \hyperref[S_11]{11})  & Nil & Nil & Nil & Inflation shock at time 30 (from 0\% to 10\% p.a.) & Varies by claim size (up to 30\% p.a. for the smallest claims, and 0 for claims exceeding \$200,000) \\ \midrule

Occurrence period superimposed inflation \hfill \hfill \linebreak (Module 8, \hyperref[S_11]{11})  & Nil & Nil & Nil & Nil & Nil for the first 20 occurrence periods; for the next 20 periods, the size of the inflation varies by claim sizes (up to 40\% reduction for the smallest claims) \\ \bottomrule
\end{tabular}
\end{table}

\subsubsection{\rev{Complexity scenarios}} \label{sssec:complexity-scenarios}

\rev{For the convenience of the user,  we provide a data generation function in \texttt{SPLICE} that outputs alternative data sets under five hypothetical scenarios ranging in data complexity. The \texttt{generate\_data()} function takes in a \texttt{complexity} parameter taking integer values between 1 (the simplest) and 5 (the most complex). Table~\ref{tab:scenarios} presents the detailed description of each scenario.}

\rev{The most complex case (5) is the default illustrated and discussed in the previous section, whereas the simplest case (1) corresponds to a chain-ladder environment as described below. The intermediate cases allow focus on particular features, and to dial complexity up or down as required.}

\rev{For the simplest case, chain-ladder compatibility is achieved in the following way:}
\begin{enumerate}[label=(\alph*)]
    \item all of \texttt{SPLICE} components Module 1 to Module 10 are defined to be independent of occurrence period; 
    \item base inflation and calendar period superimposed inflation in Module 8 and 11 occur at a constant rate per period; and
    \item occurrence period superimposed inflation in Module 8 and 11 must be independent of all other components, but otherwise can be arbitrary.
\end{enumerate}
These conditions were noted in \citet*{AvTaWaWo21} to be sufficient for homogeneity of paid losses over accident years. The time-homogeneity of the incurred loss generation, noted in Section~\ref{S_datafeatures} (Principle \ref{principles: principle14}), will then ensure that the expected distribution of incurred loss estimates across development periods is the same for all occurrence periods before the inclusion of inflation. Just as for paid losses, occurrence period SI is directly reflected in the chain ladder row parameters, which can be arbitrary. To consider the other forms of inflation, recall from the discussion in Section~\ref{ssec: treatment of inflation} that case estimates of incurred loss include base inflation to the date of estimate and calendar period SI up to settlement date.

It is already noted above that the expected distribution of incurred loss estimates across development periods is the same for all occurrence periods before the inclusion of inflation, and this will remain the case when base inflation and calendar period SI are added at a constant (though possibly different) rates.

The above demonstrates that \texttt{SPLICE} allows the user to construct very simple or very complex data sets. Naturally, a large array of intermediate cases sit in between. Full control of that level of complexity (and knowledge of its origin), allows modelers to test or illustrate specific strengths or weaknesses of their contending model, as discussed in Section \ref{sec:intro}. This idea is used, for instance, in  \citet*{McTaMi18} and \citet*[with \texttt{SynthETIC}]{AMAvTaWo21}.

\rev{We remark that the \texttt{generate\_data()} function by no means defines the limits of the complexity that can be achieved with \texttt{SPLICE}. The function is provided for the convenience of users who wish to generate (a collection of) data sets under some representative scenarios. If more complex features are required, the user is free to modify the distributional assumptions to achieve their purpose.}

\subsubsection{\rev{Comparison with chain ladder}} \label{sssec:CL-alternative-datasets}

\begin{figure}[!htbp]
    \centering
    \includegraphics[width=0.9\textwidth]{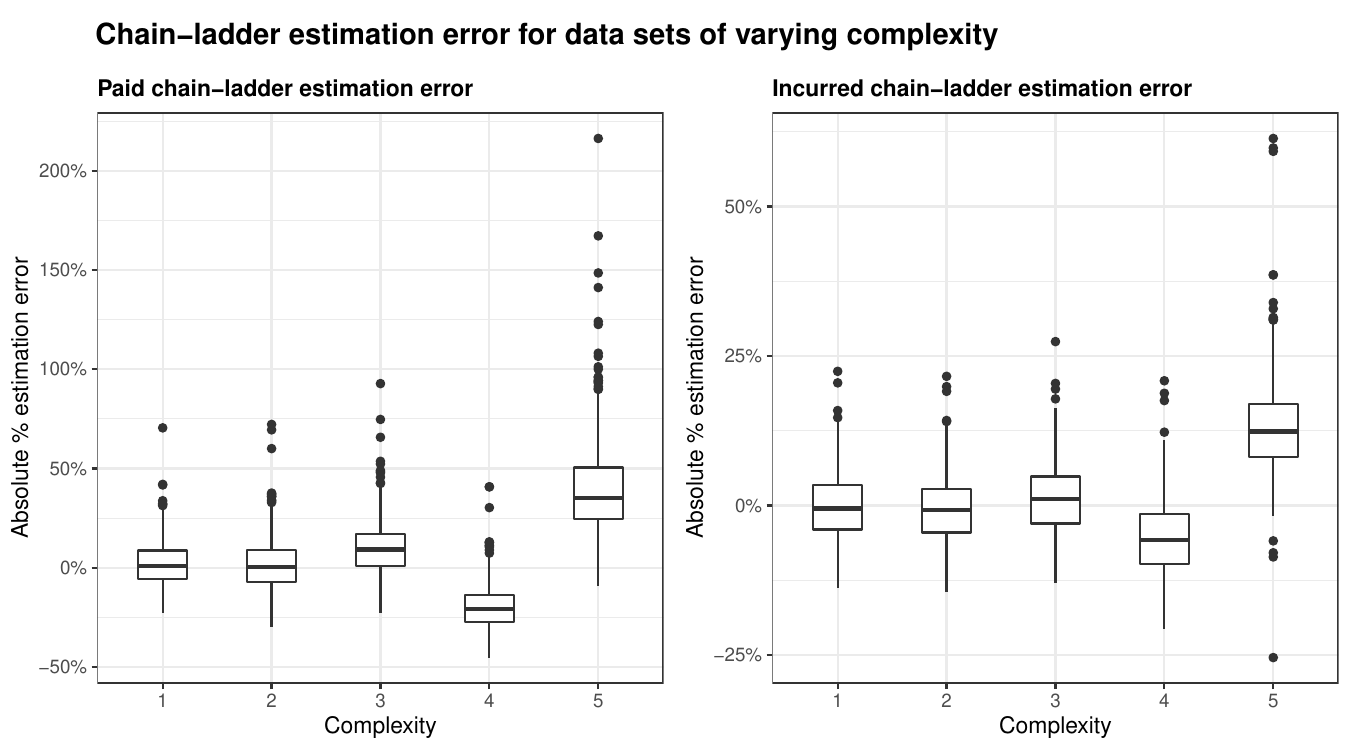}
        \caption{Chain-ladder estimation error for the five different complexity scenarios, with 500 simulations each}
        \label{fig:boxplots}
\end{figure}

\rev{Figure~\ref{fig:boxplots} plots the distribution of the chain ladder estimation error for the five scenarios described above, with 500 simulations for each scenario. The estimation error here refers to the percentage deviation of the chain ladder estimate of outstanding claim liabilities from the true value (i.e. produced by the simulator).}

\rev{The plot demonstrates that, regardless of whether we use paid (left panel) or incurred triangles (right panel) to estimate reserves, there is a consistent increase in the chain ladder estimation error as the complexity of the data set grows. This is an immediate result of the design of the data sets as outlined in Section~\ref{sssec:complexity-scenarios}: the simplest scenario is designed to provide chain ladder compatibility, whereas the most complex scenario is in complete breach of chain ladder assumptions, as visualized in Appendix~\ref{appendix:alternative-data}.}

\rev{A side observation is that the incurred chain ladder method (right panel) consistently produces significantly lower prediction errors than the paid method, which provides a further motivation for a simulator like \texttt{SPLICE} that is capable of generating case estimates. The relatively large error associated with the paid method is driven by the unstable age-to-age factors in the earlier development periods, which translate into highly variable estimates of ultimate claims for the most recent occurrence periods.}

\rev{The plot illustrates the range of complexity that can be achieved with \texttt{SPLICE}. A modeler interested in testing the effectiveness of their model in dealing with different scenarios can use the above as a starting point.}

\section{Conclusion}\label{sec:conclusion}

\texttt{SPLICE} is a CRAN claim simulator package that extends \texttt{SynthETIC} \citep*{R-SynthETIC} so that it now simulates both claim payments and estimates of incurred loss, and this enhanced version is described in the foregoing sections. Its users can access and modify its code freely, and it comes with a full set of default options which have been designed to be realistic as described in Principles~\ref{principles: principle1}--\ref{principles: principle14}, as well as [4.3.1] to [4.3.7] of \citet*{AvTaWaWo21}.

Of previously existing simulators, to the extent of the authors' knowledge, only one simulator is capable of generating sequence of case estimates of incurred losses through the lifetime of each claim. As outlined in Section~\ref{ssec: simulation lit}, it is subject to several limitations, many of which are addressed in \texttt{SPLICE}. The development of such a data generator is motivated by the scarcity of granular data with realistic features, the availability of which is essential for model development \citep*{EmWu21}.

The backward computation algorithm that \texttt{SPLICE} adopts guarantees that the final incurred estimate coincides with the claim size. This simplifies the task of generating the evolution history of case estimates to the simulation of major and minor revisions. \texttt{SPLICE} can flexibly generate the different variates of major and minor revisions. Furthermore, thanks to its modular structure, \texttt{SPLICE} allows the user to modify its default dependence structure by adjusting parameters within a module or replacing with their own. This enables the user to validate a proposed model against data of any desired level of complexity\rev{, as facilitated by the function \texttt{generate\_data()} and discussed in Section~\ref{ssec:alternative-datasets}}.

By default, \texttt{SPLICE} incorporates complex dependencies that reflect a realistic claim process (e.g. between different revisions of incurred loss estimates), producing desirable data features outlined in Section~\ref{S_datafeatures}. This may be of use in testing granular models, which sometimes include unrealistic assumptions of independence between different components. \texttt{SPLICE} may be of especial value in testing models that estimate reserves from incurred (and paid) loss data, such as the paid incurred chain reserving method of \citet*{MeWu10}.

\section*{Acknowledgements}

The authors gratefully acknowledge productive discussions on the extension of \texttt{SynthETIC} with Bernard Wong. The authors are also grateful to William Ho for research assistance, and comments that led to improvements of the R code. 

This research was supported under Australian Research Council's Linkage (LP130100723) and Discovery (DP200101859) Projects funding schemes. Melantha Wang acknowledges financial support from UNSW Australia Business School. The views expressed herein are those of the authors and are not necessarily those of the supporting organisations.

\bibliographystyle{elsarticle-harv}
\bibliography{libraries}

\begin{thebibliography}{23}
\expandafter\ifx\csname natexlab\endcsname\relax\def\natexlab#1{#1}\fi
\expandafter\ifx\csname url\endcsname\relax
  \def\url#1{\texttt{#1}}\fi
\expandafter\ifx\csname urlprefix\endcsname\relax\def\urlprefix{URL }\fi

\bibitem[{Al-Mudafer et~al.(2021)Al-Mudafer, Avanzi, Taylor, and
  Wong}]{AMAvTaWo21}
Al-Mudafer, M.~T., Avanzi, B., Taylor, G.~C., Wong, B., 2021. Stochastic loss
  reserving with mixture density neural networks. arXiv stat.ME 2108.07924.

\bibitem[{Avanzi et~al.(2021{\natexlab{a}})Avanzi, Taylor, and Wang}]{R-SPLICE}
Avanzi, B., Taylor, G., Wang, M., 2021{\natexlab{a}}. \texttt{SPLICE}:
  Synthetic paid loss and incurred cost experience (splice) simulator.
  \url{https://CRAN.R-project.org/package=SPLICE}.

\bibitem[{Avanzi et~al.(2021{\natexlab{b}})Avanzi, Taylor, Wang, and
  Wong}]{AvTaWaWo21}
Avanzi, B., Taylor, G., Wang, M., Wong, B., 2021{\natexlab{b}}.
  \texttt{SynthETIC}: An individual insurance claim simulator with feature
  control. Insurance: Mathematics and Economics 100, 296--308.

\bibitem[{Avanzi et~al.(2021{\natexlab{c}})Avanzi, Taylor, Wang, and
  Wong}]{R-SynthETIC}
Avanzi, B., Taylor, G., Wang, M., Wong, B., 2021{\natexlab{c}}.
  \texttt{SynthETIC}: Synthetic experience tracking insurance claims.
  \url{https://CRAN.R-project.org/package=SynthETIC}.

\bibitem[{Bear et~al.(2020)Bear, Shang, and You}]{R-cascsim}
Bear, R., Shang, K., You, H., 2020. \texttt{cascsim}: Casualty actuarial
  society individual claim simulator.
  \url{https://CRAN.R-project.org/package=cascsim}.

\bibitem[{{CAS Loss Simulation Model Working Party}(2007)}]{CAS07}
{CAS Loss Simulation Model Working Party}, 2007. Parameterizing the loss
  simulation model. \url{https://www.casact.org/research/lsmwp/bsupaper.pdf}.

\bibitem[{{CAS Loss Simulation Model Working Party}(2011)}]{CAS11}
{CAS Loss Simulation Model Working Party}, 2011. Modeling loss emergence and
  settlement processes. Casualty Actuarial Societly E-Forum.

\bibitem[{Cot\'e et~al.(2020)Cot\'e, Hartman, Mercier, Meyers, Cummings, and
  Harmon}]{Co20}
Cot\'e, M.-P., Hartman, B., Mercier, O., Meyers, J., Cummings, J., Harmon, E.,
  2020. Synthesizing property \& casualty ratemaking datasets using generative
  adversarial networks.

\bibitem[{De~Felice and Moriconi(2019)}]{DeMo19}
De~Felice, M., Moriconi, F., 2019. Claim watching and individual claims
  reserving using classification and regression trees. Risks 7~(4), 1--36.

\bibitem[{Dutang and Charpentier(2019)}]{R-CASdatasets}
Dutang, C., Charpentier, A., 2019. \texttt{CASdatasets}: Insurance datasets.
  \url{http://cas.uqam.ca/}.

\bibitem[{Dutang et~al.(2008)Dutang, Goulet, and Pigeon}]{R-actuar}
Dutang, C., Goulet, V., Pigeon, M., 2008. \texttt{actuar}: An \texttt{R}
  package for actuarial science. Journal of Statistical Software 25~(7), 38.

\bibitem[{Embrechts and W\"{u}thrich(2021)}]{EmWu21}
Embrechts, P., W\"{u}thrich, M.~V., 2021. Recent challenges in actuarial
  science. Annual Review of Statistics and Its Application in press.

\bibitem[{Gabrielli and W\"{u}thrich(2018)}]{GaWu18}
Gabrielli, A., W\"{u}thrich, M.~V., 2018. Individual claims history simulation
  machine. Risks 6~(2), 29.

\bibitem[{Harej et~al.(2017)Harej, G\"{a}chter, and Jamal}]{HaGaJa17}
Harej, B., G\"{a}chter, R., Jamal, S., 2017. Individual claim development with
  machine learning.
  \url{http://www.actuaries.org/ASTIN/Documents/ASTIN_ICDML_WP_Report_final.pdf}.

\bibitem[{Mack and Quarg(2004)}]{MaQu04}
Mack, T., Quarg, G., 2004. {M}unich {C}hain {L}adder. Bl{\"a}tter der DGVFM
  26~(4), 597--630.

\bibitem[{McGuire(2007)}]{McG07b}
McGuire, G., 2007. Individual claim modelling of {CTP} data. In: XIth Accident
  Compensation Seminar. Institute of Actuaries of Australia, Sydney, Australia.

\bibitem[{McGuire et~al.(2018)McGuire, Taylor, and Miller}]{McTaMi18}
McGuire, G., Taylor, G., Miller, H., 2018. Self-assembling insurance claim
  models using regularized regression and machine learning. Variance (in
  press). Also in SSRN.

\bibitem[{Merz and W{\"u}thrich(2010)}]{MeWu10}
Merz, M., W{\"u}thrich, M.~V., 2010. Paid--incurred chain claims reserving
  method. Insurance: Mathematics and Economics 46~(3), 568--579.

\bibitem[{Taylor(2019)}]{Tay19}
Taylor, G., 2019. Claim models: granular and machine learning forms. Risks
  7~(3), 82.

\bibitem[{Taylor and McGuire(2004)}]{TaMc04}
Taylor, G., McGuire, G., 2004. Loss reserving with glms: A case study. Casualty
  Actuarial Society Discussion Paper Program, Applying and Evaluating
  Generalised Linear Models.

\bibitem[{Taylor et~al.(2003)Taylor, McGuire, and Greenfield}]{TaMcGr03}
Taylor, G., McGuire, G., Greenfield, A., 2003. Loss reserving: {Past}, present
  and future, university of Melbourne Research Paper.

\bibitem[{Taylor et~al.(2008)Taylor, McGuire, and Sullivan}]{TaMcSu08}
Taylor, G., McGuire, G., Sullivan, J., 2008. Individual claim loss reserving
  conditioned by case estimates. Annals of Actuarial Science 3~(1-2), 215--256.

\bibitem[{Teugels and Sundt(2004)}]{TaSu04}
Teugels, J.~L., Sundt, B. (Eds.), 2004. Encyclopedia of Actuarial Science. John
  Wiley \& Sons, Ltd.

\end{thebibliography}

\newpage

\begin{appendices}

\section{Parametrizations}\label{appendix:A}

The following table displays the formal parameterization of Modules 9 to 11 for the example of Section~\ref{ssec:example-implementation}.

\begin{center}

\small

\begin{longtable}{llp{3.5cm}P{3.5cm}p{7.5cm}}
        \toprule
        \multicolumn{3}{c}{\textbf{Parameter Type}} & \textbf{Functional Form} & \textbf{Numerical Parameters} \\
        
        \midrule
        \multicolumn{3}{l}{\textbf{Global}} & & Time unit $= 1/4$ (Calendar quarter) \\
        & & & & Reference claim size $= 200,000$ \\
        
        \midrule
        \multicolumn{3}{p{4cm}}{\textbf{Major revisions}} & & \\
        & \multicolumn{2}{p{4cm}}{\textbf{Number of revisions of incurred loss}} & $F_{K|s}^{\text{Ma}}(k;s)$ & For $s_{ir} \leq 15000$, \quad $\mathbb{P}(K_{ir}^{\text{Ma}} > 1) = 0$\\
        & & & & For $s_{ir} > 15000$: 
        \begin{itemize}
            \item[] $\mathbb{P}(K_{ir}^{\text{Ma}} = 2) = 0.1 + 0.3 \min(1, \frac{s_{ir} - 15000}{185000});$
            \item[] $\mathbb{P}(K_{ir}^{\text{Ma}} = 3) = 0.5 \min(1, \frac{\max(0, s_{ir} - 15000)}{185000});$
            \item[] $\mathbb{P}(K_{ir}^{\text{Ma}} > 3) = 0$
        \end{itemize} \\ \addlinespace
        
        & \multicolumn{2}{p{4cm}}{\textbf{Epochs of revisions of incurred loss}} & $\mathbb{P}(\tau_{irl}^{\text{Ma}} = w_{ir}^{(m_{ir}-1)}|s_{ir})$ for $l=k_{ir}^{\text{Ma}}$ & Major revisions occur only for $m_{ir} \geq 4$ \\
        & & & & ${\mathbb{P}(\tau_{irl}^{\text{Ma}} = w_{ir}^{(m_{ir}-1)}|s_{ir})=}$ $0.2\min \left( 1, \frac{\max(0, s_{ir} - 200000)}{2800000} \right)$ \\ \addlinespace
        & & & $F_{T|w}^{\text{Ma}1}(\tau;w)$ & The $\tau_{irl}^{\text{Ma}}$, $l=2,\dots,k_{ir}^{\text{Ma}}-1$ are sampled from a triangular pdf with range $\left( w_{ir}^{(m_{ir}-1)}⁄3, w_{ir}^{(m_{ir}-1)} \right)$  and maximum density at $w_{ir}^{(m_{ir}-1)}⁄3$  \\ \addlinespace
        & & & $F_{T|w}^{\text{Ma}2}(\tau;w)$ & The $\tau_{irl}^{\text{Ma}}$, $l=2,\dots, k_{ir}^{\text{Ma}}$ are sampled from a triangular pdf with range $\left( w_{ir}⁄3, w_{ir} \right)$  and maximum density at $w_{ir}⁄3$ \\
        
        & \multicolumn{2}{p{4cm}}{\textbf{Revision factors (multipliers of incurred losses)}} & $F_{G|h}^{\text{Ma}}(g;h)$ & For $l=2$, $\ln{g_{irl}^{\text{Ma}}}$ drawn from $\mathcal{N}(1.8, 0.2^2)$ \\
        & & & & For $l=3$, $\ln{g_{irl}^{\text{Ma}}}$ drawn from $\mathcal{N}(\mu, 0.2^2)$ with $\mu = 1 + 0.07(6 - g_{ir2}^{\text{Ma}})$ \\
        & & & & In each case $g_{irl}^{\text{Ma}}$ is subject to an upper bound of $\frac{0.95y_{ir}(\tau)}{c_{ir}(\rev{\tau^{-}})}$, where the revision is assumed to occur with delay $\tau$ from notification \\ \addlinespace

        \midrule
        \multicolumn{3}{p{4cm}}{\textbf{Minor revisions}} & & \\
        
        & \multicolumn{2}{p{4cm}}{\textbf{Number of revisions of incurred loss}} & & \\
        & & \textbf{Simultaneous with partial payment} & $F_{B|l}(b)$ & $F_{B|l}(0)$ = $F_{B|l}(1) = \tfrac{1}{2}$ \\ \addlinespace
        
        & & \textbf{Not simultaneous with partial payment} & $F_{K|w}^{\text{Mi}2}(k;w)$ & Distribution of $K_{ir}^{\text{Mi}2}$ is geometric, with mean $= \min(3, \frac{w_{ir}}{4})$ \\ \addlinespace
        
        & \multicolumn{2}{p{4cm}}{\textbf{Epochs of revisions of incurred loss}} & & \\
        & & \textbf{Simultaneous with partial payment} & & Coincide with the partial payments\\
        
        & & \textbf{Not simultaneous with partial payment} & $F_{T|w}^{\text{Mi}}(\tau;w) $ & The $\tau_{irl}^{\text{Mi}}$, $l=1,\dots,k_{ir}^{\text{Mi}}$ are drawn from a uniform distribution with range $\left( w_{ir}⁄6, w_{ir} \right)$\\
        
        & \multicolumn{2}{p{4cm}}{\textbf{Revision factors (multiplier of outstanding paid losses)}} & $F_{G|w, \tau}^{\text{Mi}}(g;w, \tau) $ & For $\tau_{irl}^{\text{Mi}} \leq \frac{w_{ir}}{3}$, $\ln{g_{irl}^{\text{Mi}}}$ is drawn from:
        \begin{itemize}
            \item[] $\mathcal{N}(0.15, 0.05^2)$ if preceded by a 2\textsuperscript{nd} major revision (i.e. subsequent to notification);
            \item[] $\mathcal{N}(0.15, 0.1^2)$ otherwise
        \end{itemize} \\
        & & & & For $\frac{w_{ir}}{3} < \tau_{irl}^{\text{Mi}} \leq \frac{2w_{ir}}{3}$, $\ln{g_{irl}^{\text{Mi}}}$ is drawn from:
        \begin{itemize}
            \item[] $\mathcal{N}(0, 0.05^2)$ if preceded by a 2\textsuperscript{nd} major revision;
            \item[] $\mathcal{N}(0, 0.1^2)$ otherwise
        \end{itemize} \\
        & & & & For $\tau_{irl}^{\text{Mi}} < \frac{2w_{ir}}{3}$, $\ln{g_{irl}^{\text{Mi}}}$ is drawn from:
        \begin{itemize}
            \item[] $\mathcal{N}(-0.1, 0.05^2)$ if preceded by a 2\textsuperscript{nd} major revision;
            \item[] $\mathcal{N}(-0.1, 0.1^2)$ otherwise
        \end{itemize} \\
        & & & & In each case, $g_{irl}^{\text{Mi}}$ is subject to the constraint that $y_{ir}(\rev{\tau^{-}}) = c_{ir}(\rev{\tau^{-}}) + x_{ir}(\rev{\tau^{-}}) \geq kc_{ir}(\rev{\tau^{-}})$, or equivalently $x_{ir} (\rev{\tau^{-}})=(y_{ir} (\tau)-c_{ir} (\rev{\tau^{-}}))⁄g_{irl}^{\text{Mi}} \geq (k-1)c_{ir} (\rev{\tau^{-}})$, where the revision is assumed to occur with delay $\tau$ from notification and $k>1$ is a fixed constant (not necessarily the same $k$ as above). The current program assumes $1⁄k=0.95$ \\
        
        \midrule
        \multicolumn{3}{l}{\textbf{Base inflation}} & $f(\bar{t})$ & $=(1+\alpha)^{\bar{t}}$, where $\alpha$ is equivalent to an increase of 2\% p.a., allowing for the relevant time units \\
        
        \midrule
        \multicolumn{3}{p{3cm}}{\textbf{Superimposed inflation}} & $g_O (u|s)$ & $=1$ for $u \leq 20$ \\
        & & & & $= 1 - 0.4\max \left( 0, 1 - \frac{s}{50000} \right)$ for $u > 20$ \\ \addlinespace
        & & & $g_C(\bar{t}|s)$ & $= \left( 1 + \beta(s) \right)^{\bar{t}}$ with $\beta(s) = \gamma \max \left( 0, 1 - \frac{s}{200000} \right) $ and $\gamma$ is equivalent to a 30\% p.a. inflation rate, allowing for the relevant time units \\
        \bottomrule
\end{longtable}
\label{tab: formal parameterizations}
\end{center}
{\footnotesize Notes about claim sizes and inflation:
    \begin{enumerate}[label=(\alph*)]
        \item Some components are defined in terms of claim size.  The definition then displays claim size in raw uninflated units. The inputs to the example application of \texttt{SPLICE}, on the other hand, express claim sizes as multiples of a reference claim size equal to 200,000.  For example, the amount of 15,000 that appears in the definition of the frequency of major revisions is expressed in \texttt{SPLICE} as $0.075 \times 200,000$.
        \item Wherever claim size ($s$ or $s_{ir}$) is compared with a numerical quantity (e.g. 50,000 for SI), the claim size \textbf{excludes} all forms of inflation.
    \end{enumerate}
}

\newpage
\section{Cumulative Incurred Loss Triangle of the Example Implementation}\label{appendix:triangles}
\vspace{-2mm}
For space considerations, below we show only the claim development triangles on a \textbf{yearly} basis (with the inclusion of inflation); however, the underlying data is calculated based on quarterly development pattern and is available on the \texttt{SPLICE} repository (see Section~\ref{sec:SPLICE repo}).

\begin{table}[htbp!]
    \centering
    \footnotesize
    \begin{tabular}{crrrrrrrrrr}
    \toprule
    & \multicolumn{10}{c}{Development Year} \\ \midrule
    Accident Year & 1      & 2      & 3      & 4      & 5      & 6      & 7     & 8     & 9     & 10    \\
    \midrule
    1 & 16,389 & 39,274 & 52,490 & 64,217 & 66,874 & 73,648 & 76,208 & 77,264 & 77,173 & \textbf{77,211} \\
    2 & 18,017 & 42,795 & 58,207 & 67,869 & 74,310 & 75,871 & 78,755 & 80,789 & \textbf{80,997} & 81,360 \\
    3 & 21,714 & 52,799 & 65,190 & 71,618 & 78,491 & 81,707 & 82,539 & \textbf{83,320} & 83,431 & 83,230 \\
    4 & 21,172 & 48,529 & 64,916 & 75,795 & 83,538 & 87,737 & \textbf{87,446} & 87,523 & 87,610 & 86,801 \\
    5 & 26,459 & 64,570 & 80,520 & 94,317 & 98,718 & \textbf{100,446} & 102,592 & 103,159 & 102,880 & 103,254 \\
    6 & 29,379 & 71,254 & 90,285 & 100,748 & \textbf{104,567} & 106,252 & 108,053 & 107,613 & 107,433 & 107,352 \\
    7 & 21,154 & 64,467 & 82,384 & \textbf{97,824} & 103,947 & 107,824 & 108,037 & 107,518 & 107,332 & 107,334 \\
    8 & 38,244 & 82,992 & \textbf{104,322} & 114,536 & 117,436 & 120,199 & 122,193 & 124,254 & 125,922 & 125,889 \\
    9 & 34,735 & \textbf{97,448} & 125,597 & 140,065 & 146,810 & 148,883 & 151,319 & 152,575 & 152,411 & 152,916 \\
    10 & \textbf{49,339} & 120,973 & 149,794 & 160,851 & 167,556 & 167,292 & 166,790 & 166,531 & 166,428 & 166,216\\ \bottomrule
    \end{tabular}
    \caption{Cumulative incurred loss triangle of simulated claims (\$000)}
    \label{tab:simulated triangle}
\end{table}

\begin{table}[htbp!]
    \centering
    \footnotesize
    \begin{tabular}{crrrrrrrrrr}
    \toprule
    & \multicolumn{10}{c}{Development Year} \\ \midrule
    Accident Year & 1      & 2      & 3      & 4      & 5      & 6      & 7     & 8     & 9     & 10    \\
    \midrule
    1 & 16,389 & 39,274 & 52,490 & 64,217 & 66,874 & 73,648 & 76,208 & 77,264 & 77,173 & \textbf{77,211} \\
    2 & 18,017 & 42,795 & 58,207 & 67,869 & 74,310 & 75,871 & 78,755 & 80,789 & \textbf{80,997} & 81,037 \\
    3 & 21,714 & 52,799 & 65,190 & 71,618 & 78,491 & 81,707 & 82,539 & \textbf{83,320} & 83,382 & 83,424 \\
    4 & 21,172 & 48,529 & 64,916 & 75,795 & 83,538 & 87,737 & \textbf{87,446} & 88,871 & 88,938 & 88,982 \\
    5 & 26,459 & 64,570 & 80,520 & 94,317 & 98,718 & \textbf{100,446} & 102,331 & 103,998 & 104,076 & 104,127 \\
    6 & 29,379 & 71,254 & 90,285 & 100,748 & \textbf{104,567} & 109,114 & 111,162 & 112,973 & 113,058 & 113,113 \\
    7 & 21,154 & 64,467 & 82,384 & \textbf{97,824} & 104,406 & 108,946 & 110,991 & 112,800 & 112,884 & 112,940 \\
    8 & 38,244 & 82,992 & \textbf{104,322} & 120,878 & 129,012 & 134,622 & 137,148 & 139,383 & 139,487 & 139,556 \\
    9 & 34,735 & \textbf{97,448} & 124,934 & 144,761 & 154,502 & 161,221 & 164,246 & 166,923 & 167,048 & 167,130 \\
    10 & \textbf{49,339} & 122,472 & 157,017 & 181,935 & 194,177 & 202,621 & 206,423 & 209,787 & 209,944 & 210,048\\ \bottomrule
    \end{tabular}
    \caption{Cumulative claim development triangle predicted by the chain ladder model (\$000)}
    \label{tab:CL triangle}
\end{table}

The code to produce the above triangles builds on the sample code in Section~\ref{sssec:example-code} and is provided below:
\begin{lstlisting}[language=R]
## SPLICE simulated incurred loss triangle
incurred_inflated <- output_incurred(
  test_inflated, incremental = F, aggregate_level = 4)

## Chain-ladder predicted incurred loss
# output the past cumulative incurred triangle simulated by SPLICE
cumtri <- output_incurred(
  test_inflated, aggregate_level = 4, incremental = F, future = F)
# calculate the age to age factors
selected <- attr(ChainLadder::ata(cumtri), "vwtd")
# complete the triangle
CL_prediction <- cumtri
J <- nrow(cumtri)
for (i in 2:J) {
  for (j in (J - i + 2):J) {
    CL_prediction[i, j] <- CL_prediction[i, j - 1] * selected[j - 1]
  }
}
\end{lstlisting}

\newpage
\section{Excerpt of Simulated Data}\label{appendix:data excerpt}
Table~\ref{tab:data excerpt} is an excerpt of the transactional dataset generated by \texttt{SPLICE} (prior to the addition of inflation) and displays the full claim history of claims \#2 and \#40 (in Figure~\ref{fig:example claim history}) from notification to settlement. Table~\ref{tab:data description} provides the detailed description of the variables.

The equations~\eqref{eqn: sir} to \eqref{eqn: yir constraint 2}, which describe the computation of case estimates before the effect of inflation is added, may be illustrated by a numerical example. The third transaction record for claim \#2 indicates a simultaneous minor revision at the time of a partial payment, by a factor of 1.0503. The total incurred loss after this transaction is thus the sum of the revised outstanding paid loss ($21,688 \times 1.0503$) and the cumulative paid loss prior to the revision (\$2,005), as governed by Equation~\eqref{eqn: yir mi 2}. The partial payment made brings the outstanding claim liability down to \$20,654.

Major revisions are rarer, but usually of a greater magnitude. Claim \#40 sees a major revision 9.824 periods after notification (at transaction time 14.102). The major revision effects a change on the incurred loss directly ($52,\!969 \times 3.1759$), driving up the estimated incurred loss and outstanding payments to \$168,224 and \$148,497 respectively.

We remark that in practice, the computation algorithm simulates the case estimates in reverse chronological order to ensure that the final incurred estimate coincides with the claim size; see Section~\ref{sec:architecture}.

The code to generate this data set is provided in Section~\ref{sssec:example-code}.

\begin{table}[!htbp]
\centering
\footnotesize
\begin{tabular}{@{}rrrrrrrrr@{}}
\toprule
\verb=claim_no= & \verb=claim_size= & \verb=txn_time= & \verb=txn_delay= & \verb=txn_type= & \verb=incurred= & \verb=OCL= & \verb=cumpaid= & \verb=multiplier= \\ \midrule
2 & 22,562 & 1.298 & 0.000 & Ma & 21,635 & 21,635 & 0 & 1.0000 \\
2 & 22,562 & 2.203 & 0.906 & PMi & 23,694 & 21,688 & 2,005 & 1.0952 \\
2 & 22,562 & 2.695 & 1.397 & PMi & 24,784 & 20,654 & 4,130 & 1.0503 \\
2 & 22,562 & 3.317 & 2.019 & PMi & 22,562 & 2,446 & 20,116 & 0.8924 \\
2 & 22,562 & 3.629 & 2.332 & P & 22,562 & 0 & 22,562 &  \\ \midrule
40 & 143,183 & 4.278 & 0.000 & Ma & 15,969 & 15,969 & 0 & 1.0000 \\
40 & 143,183 & 4.456 & 0.177 & P & 15,969 & 13,901 & 2,068 &  \\
40 & 143,183 & 6.062 & 1.784 & P & 15,969 & 11,957 & 4,013 &  \\
40 & 143,183 & 7.335 & 3.056 & P & 15,969 & 9,296 & 6,674 &  \\
40 & 143,183 & 7.503 & 3.225 & Mi & 17,402 & 10,728 & 6,674 & 1.1541 \\
40 & 143,183 & 7.537 & 3.258 & Mi & 18,318 & 11,644 & 6,674 & 1.0854 \\
40 & 143,183 & 8.177 & 3.899 & P & 18,318 & 9,846 & 8,472 &  \\
40 & 143,183 & 9.116 & 4.837 & Mi & 19,218 & 10,745 & 8,472 & 1.0913 \\
40 & 143,183 & 9.121 & 4.842 & P & 19,218 & 8,972 & 10,245 &  \\
40 & 143,183 & 10.162 & 5.884 & P & 19,218 & 6,929 & 12,289 &  \\
40 & 143,183 & 10.480 & 6.202 & P & 19,218 & 5,141 & 14,077 &  \\
40 & 143,183 & 11.133 & 6.855 & PMi & 18,769 & 2,891 & 15,878 & 0.9127 \\
40 & 143,183 & 12.017 & 7.738 & Mi & 18,918 & 3,040 & 15,878 & 1.0516 \\
40 & 143,183 & 12.442 & 8.163 & PMi & 18,595 & 930 & 17,665 & 0.8938 \\
40 & 143,183 & 12.495 & 8.216 & Mi & 18,595 & 930 & 17,665 & 1.0149 \\
40 & 143,183 & 12.568 & 8.290 & Ma & 53,754 & 36,089 & 17,665 & 6.1785 \\
40 & 143,183 & 13.623 & 9.345 & Mi & 54,116 & 36,451 & 17,665 & 1.0100 \\
40 & 143,183 & 13.697 & 9.418 & PMi & 52,969 & 33,242 & 19,727 & 0.9685 \\
40 & 143,183 & 14.102 & 9.824 & Ma & 168,224 & 148,497 & 19,727 & 3.1759 \\
40 & 143,183 & 14.109 & 9.830 & P & 168,224 & 146,744 & 21,480 &  \\
40 & 143,183 & 14.925 & 10.646 & Mi & 164,702 & 143,222 & 21,480 & 0.9760 \\
40 & 143,183 & 15.402 & 11.123 & Mi & 149,735 & 128,255 & 21,480 & 0.8955 \\
40 & 143,183 & 15.589 & 11.310 & P & 149,735 & 126,591 & 23,145 &  \\
40 & 143,183 & 17.114 & 12.836 & PMi & 147,069 & 17,189 & 129,880 & 0.9789 \\
40 & 143,183 & 17.414 & 13.136 & Mi & 145,194 & 15,314 & 129,880 & 0.8909 \\
40 & 143,183 & 18.334 & 14.055 & PMi & 143,183 & 0 & 143,183 & 0.8687 \\ \bottomrule
\end{tabular}
\caption{Transaction records of two sample claims from the example implementation}
\label{tab:data excerpt}
\end{table}

\begin{table}[!htbp]
\centering
\footnotesize
\begin{tabular}{p{2cm}p{13.73cm}}
\toprule
Variable & Description \\ \midrule
\verb=claim_no= & claim number, which uniquely characterises each claim. \\
\verb=claim_size= & size of the claim (in constant dollar values). \\
\verb=txn_time= & the calendar time of transaction (on a continuous time scale, with origin at the beginning of occurrence period 1). \\
\verb=txn_delay= & delay from notification to the subject transaction. \\
\verb=txn_type= & nature of the transactions, ``Ma" for major revision, ``Mi" for minor revision, ``P" for payment, ``PMa" for major revision coincident with a payment, ``PMi" for minor revision coincident with a payment. \\
\verb=incurred= & case estimate of total incurred loss immediately after the transaction. \\
\verb=OCL= & case estimate of outstanding claim payments immediately after the transaction. \\
\verb=cumpaid= & cumulative claim paid after the transaction. \\
\verb=multiplier= & revision multipliers (subject to further constraints specified by Equations \eqref{eqn: yir constraint} and \eqref{eqn: yir constraint 2}), missing for transactions that do not involve a revision. \\ \bottomrule
\end{tabular}
\caption{Variable descriptions}
\label{tab:data description}
\end{table}

\newpage
\section{\rev{Visualizations of Claim Development under Different Complexity Scenarios}}\label{appendix:alternative-data}
\rev{For space considerations, we show only the plots for the complexity level of 1 and 5. At a complexity level of 1, the clustered lines across all occurrence periods in both plots indicate the homogeneity of claim payments and incurred losses across all occurrence periods. This contrasts with the two plots in the bottom featuring a steady shortening of both paid and incurred patterns with respect to the occurrence period, which makes the modeling a challenge.}

\begin{figure}[!ht]
    \centering
    \includegraphics[width=\textwidth]{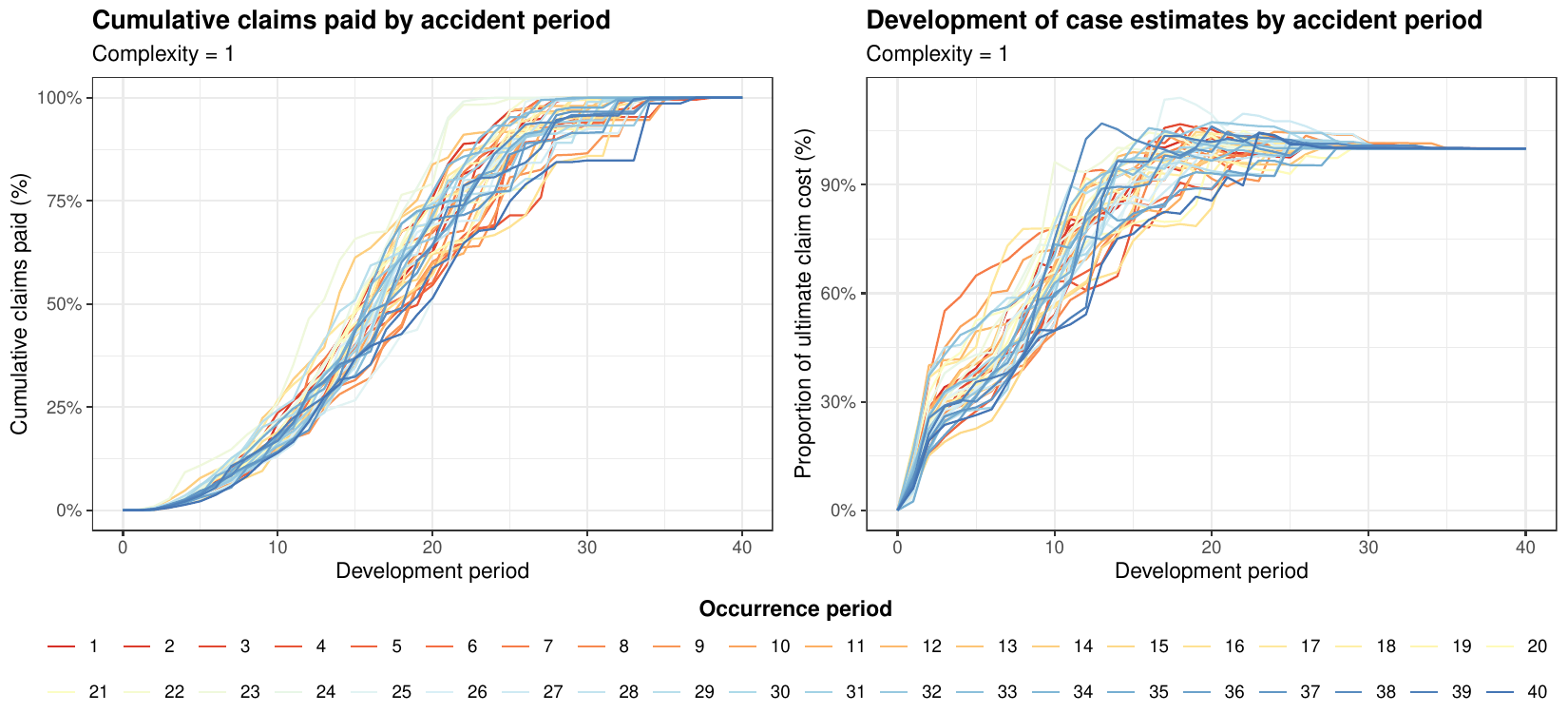}
    \includegraphics[width=\textwidth]{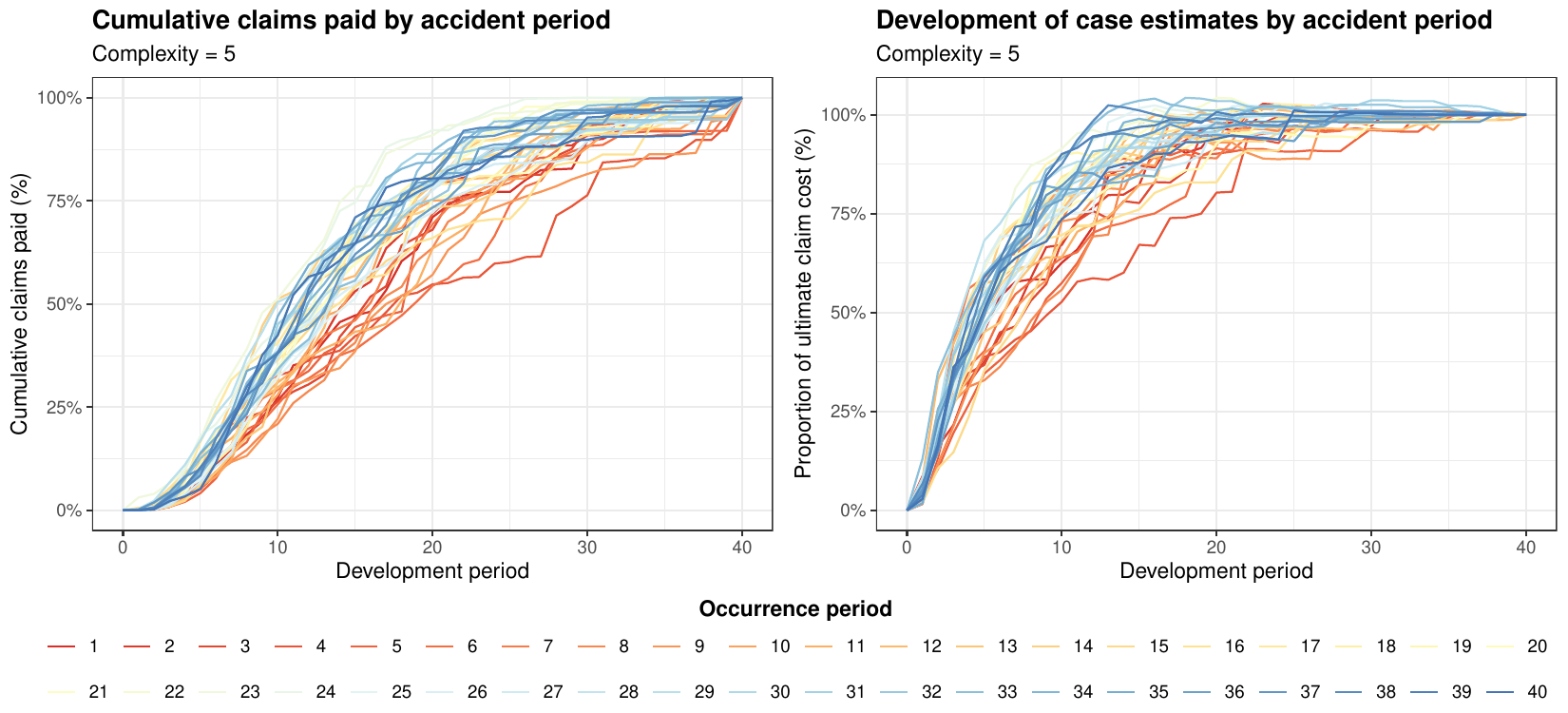}
        \caption{Claim development pattern under different complexity scenarios (1 represents the simplest, 5 represents the most complex, in our case, the default portfolio described in Section~\ref{ssec:example-implementation})}
        \label{fig:rfun}
\end{figure}

\end{appendices}

\end{document}